\documentclass[%
reprint,
 superscriptaddress,
 amsmath,amssymb,
 aps,
 nofootinbib,
 prb,
 physics
]{revtex4-2}

\usepackage{graphicx} % Required for inserting images
\usepackage{subcaption}
\usepackage{xcolor}
\usepackage{amsmath}
\usepackage{amssymb}
\usepackage{amscd}
\usepackage{dsfont}
\usepackage{enumerate}
\usepackage{enumitem}
\usepackage{amsfonts}
\usepackage{epsfig}
\usepackage{esint}
\usepackage{booktabs}
\usepackage{array}
\usepackage{color}
\usepackage{mathtools}
\usepackage{yfonts}
\usepackage{bbold}
\usepackage[utf8]{inputenc}
\usepackage[english]{babel}
\usepackage{import}
\usepackage{caption}
\usepackage{multirow}
\usepackage{cases}

\def\eqa{\begin{eqnarray}}
\def\eqae{\end{eqnarray}}
\def\eq{\begin{equation}}
\def\eqe{\end{equation}}
\def\be{\begin{equation}}
\def\ee{\end{equation}}
\def\bea{\begin{eqnarray}}
\def\eea{\end{eqnarray}}
\def\ba{\begin{array}}
\def\ea{\end{array}}

\begin{document}

\author{Yongjiang Xu}
\affiliation{School of Quantum \& Kavli Institute of Theoretical Sciences, University of Chinese Academy of Sciences, Beijing 100190, China}

\author{Weixin Sun}
\affiliation{School of Quantum \& Kavli Institute of Theoretical Sciences, University of Chinese Academy of Sciences, Beijing 100190, China}

\author{Chushun Tian}
\affiliation{Institute of Theoretical Physics, Chinese Academy of Sciences, Beijing 100190, China}

\author{Huajia Wang}
\email{wanghuajia@ucas.ac.cn}
\affiliation{School of Quantum \& Kavli Institute of Theoretical Sciences, University of Chinese Academy of Sciences, Beijing 100190, China}%Lines %break automatically or can be forced with \\
\affiliation{Peng Huanwu Center for Fundamental Theory, Hefei, Anhui 230026, China}

\date{\today}

\title{Phase transition from eigenstate thermalization: forbidden singularity and instanton proliferation via AGT correspondence}

\begin{abstract}
In theoretical physics, finding connections between problems that appear in distinct contexts is an important way to leapfrog progresses, often by illuminating deep aspects that may otherwise seem obscure. In this paper, we consider in 2d CFTs the phenomenon of forbidden singularities in auto-correlation functions -- a key signature of  eigenstate thermalization. We show that they correspond to phase transitions in the context of eigenstates. The connection is made explicit by utilizing the AGT correspondence, which relates eigenstate auto-correlations to the Nekrasov partition functions describing an instanton gas of the $\mathcal{N}=2$ SUSY gauge theories. We show that by taking the counter-part of the heavy-light limit, two phases emerge for the instanton gas. They are dominated by configurations represented by string-like Young tableaux with distinct structures and thermodynamic properties, which bare resemblance to the confined and the deconfined phases. A phase transition occurs as instantons proliferate from one side, in a manner that mimics the Lee-Yang theory. We work out the critical fugacity and find it corresponding exactly to the forbidden singularity. 

\end{abstract}

\maketitle

\section{Introduction}
The phenomenon of quantum thermalization in isolated systems is an important subject that plays key roles in many areas, including the back hole information paradox \cite{Hawking:1975IP,Hawking1976} -- especially in the context of AdS/CFT.  At its core is the conflict between the apparent thermal equilibrium at late times and the underlying unitary dynamics of pure states. An important progress in understanding the mechanism of dynamical thermalization is the eigenstate thermalization hypothesis (ETH) \cite{Srednicki:1994,Deutsch:1991,Rigol:2008,Alessio:2016}. It proposes that in the thermodynamic limit, generic high energy eigenstates $|E\rangle$, i.e. those having finite energy densities, behave like the corresponding thermal equilibrium (e.g. micro-canonical ensemble at the same energy) when probed by simple observables $\mathcal{O}_{\text{obs}}$: 
\be\label{eq:ETH_1} 
\langle E | \mathcal{O}_{\text{obs}}|E\rangle \approx \text{Tr} \left( \rho^E_{\text{micro}}\mathcal{O}_{\text{obs}}\right) 
\ee
The more complete version of ETH describes the structure for the matrix elements of $ \mathcal{O}_{\text{obs}}$ between high energy eigenstates. An alternative formulation of ETH based on subsystems was proposed in \cite{Dymarsky:2018}.

For CFTs defined on spheres, the  Hamiltonian is related to the symmetry generator under scaling. Due to the state-operator correspondence, energy eigenstates are more tractable, and properties of energy eigenstates such as ETH can be studied in more explicit terms. This is especially the case in two dimensional space-time, where the conformal symmetry is extended to the infinite dimensional virasoro algebra and thus imposing more constraints on the dynamics. As a result, 2d CFTs serve as a good theoretical context to study and clarify various aspects of ETH. For example, the nature of thermalization 2d CFTs exhibits special features due to the infinitely many additional conserved (KdV) charges \cite{Bazhanov:1994KdV,Bazhanov:1996KdV,Bazhanov:1998KdV} from the virasoro algebra that significantly constrains the pattern of thermalization. In particular, thermalization towards equilibrium is controlled by the generalized Gibbs ensembles \cite{Cardy:2016GGE} for the KdV charges. The implications on the behavior of subsystem ETH in 2d CFTs was studied in \cite{Chen:2024lji,Chen:2024ysb}. 

An interesting consequence of ETH arises when the observable takes the form of bi-local operators separated along the imaginary time domain, and consisting of light primary fields $O_L$: 
\be
\mathcal{O}_{\text{obs}} = O_L(0)O_L(\tau)
\ee
The expectation $f_E(\tau)=\langle E| \mathcal{O}_{\text{obs}}|E\rangle$ defines the auto-correlation function in the primary eigen-state $|E\rangle=O_H(-\infty)|\Omega\rangle$ on a circle of circumference $2\pi$. It is given by the Euclidean correlation function: 
\bea \label{eq:4pt_1}
f_E(\tau) &=& \langle O_H(-i\infty) O_L(0) O_L(\tau) O_H(i\infty)\rangle_{\text{cylinder}} \nonumber\\
&\sim & \langle O_L(0) O_L(z) O_H(1) O_H(\infty)\rangle_{\text{plane}}  
\eea
where we have applied the conformal map $z=1-e^{\tau+ix}$ from the cylinder to the plane. On the other hand, when ETH is valid, this expectation value should be approximated by a thermal correlation function at an appropriate temperature $\beta_H$ that is set to match the energy density of $|E\rangle$: 
\be \label{eq:ETH_3}
f_E(\tau) \approx \langle O_L(0) O_L(\tau)\rangle_{\beta_H}
\ee
The RHS of (\ref{eq:ETH_3}) as a function of $\tau$ contains an infinite number of singularities located on the imaginary time at $\tau_n=i\beta_H n,\;\;n\in \mathds{Z}$. They correspond to the OPE singularity at $\tau=0$ and its thermal images under the KMS periodicity of $\langle... \rangle_{\beta_H}$. At face value, (\ref{eq:ETH_3}) and (\ref{eq:4pt_1}) then implies that the correlation function $C(z)=\langle O_L(0)O_L(z) O_H(1) O_H(\infty) \rangle$ develops singularities at $z_n=1-e^{-\tau_n}$. This is of course an artifact of taking the leading order in the thermodynamic limit supporting (\ref{eq:ETH_3}). The only true singularities of $C(z)$ are of the OPE type located at $z=\lbrace 0,1,\infty\rbrace$, the other singularities at $z_{n\neq 0}$ should be understood as emergent phenomena in the thermodynamic limit, and are termed as ``forbidden singularities". 

In CFTs, we can use the large central charge $c$ to effectuate an alternative thermodynamic limit. This corresponds to a macroscopic number of local degrees of freedom, in contrast to a large system size in conventional thermodynamic limits. In this limit, the conformal dimension $h_H$ of $O_H$ scales with $c$ at fixed ratio $6h_H/c = \epsilon_H>1/4$, such that it creates a black hole micro-state; while that of $O_L$ does not, i.e. $h_L\sim \mathcal{O}(1)$, such that it only probes the background created by $O_H$. This is called the heavy-light limit for the correlation function. When restricted to holographic CFTs exhibiting large spectral gaps, the auto-correlation function is dominated through (\ref{eq:4pt_1}) by the $t$-channel vacuum block $\mathcal{V}^t_{vac}(z)$ in its virasoro block decomposition: 
\be 
f_E(\tau) \propto \sum_{h,\bar{h}} C^{t}_{h} \;C^{t}_{\bar{h}}\; \mathcal{V}^{t}_{h}(z) \;\mathcal{V}^{t}_{\bar{h}}(\bar{z}) \sim \mathcal{V}^{t}_{vac}(z) \mathcal{V}^{t}_{vac}(\bar{z}) 
\ee
where the superscript $t$ denotes OPE channel $HH \to LL$. As a result, the phenomenon of forbidden singularities can be confined into the large $c$ vacuum virasoro block, a highly universal kinematic object. Many studies on virasoro blocks have been carried out, it was found that forbidden singularities indeed show up in ways that are consistent with ETH, and their resolutions have been discussed in various contexts \cite{Fitzpatrick:2014,Fitzpatrick:2015, Fitzpatrick:2016ive, Fitzpatrick:2016mjq, Chen:2017,Wang:2018,Collier:2018exn}. The phenomenon of forbidden singularities constitutes a key signature of the thermalization properties of the underlying theory. As a sharp manifestation of the conflict between thermal behavior (KMS periodicity) and unitarity (only true singularities are OPE), it shares a common symptom with the black hole information paradox, so in some sense can be viewed as a baby version of the latter.  

Apart from being an intriguing mathematical fact regarding virasoro blocks, the emergence of forbidden singularities may in itself represent a broader class of phenomena that are relevant in other physical contexts. In particular, as a class of non-analyticity that appears only after taking the large $c$ limit, they are analogous to how phase transitions are characterized in thermodynamics. In this case, the non-analytic object is not a partition function but the eigenstate auto-correlation function, so it does not represent a thermodynamic phase transition, but a phase transition in isolated systems. In general, the emergence of local thermodynamics proposed by ETH naturally leads to questions concerning phase transitions in eigenstates. A major difficulty is that while a thermalized eigenstate is only well-defined in a finite (and large) system, the occurrence of phase transitions relies on strictly taking the thermodynamic limit. Fortunately, taking the large $c$ limit in CFTs provides an scenario where eigenstates remain well-defined and trackable in the $c\to \infty$ limit, thereby bypassing the aforementioned difficulty. Motivated by this, we are interested in finding connections between the forbidden singularities and phase transitions in eigenstate thermalized systems. A useful tool for this task is provided by so-called AGT correspondence \cite{Alday:2009aq,Nekrasov:2002qd,Nekrasov:2003rj,LeFloch:2020uop} that relates virasoro blocks in 2d CFTs and instanton partition functions in four dimensional $\mathcal{N}=2$ SYM gauge theories of gauge group $\text{SU}(2)$. It serves as a bridge to recast phenomena regarding virasoro blocks in a thermodynamic context, i.e. described by a partition function, of the instanton gas in 4d gauge theories. In this paper, we study the emergence of forbidden singularities as phase transitions in regarding instantons in 4d gauge theories. By doing this, we hope to probe deep into the mechanism of eigenstate thermalization in relation to forbidden singularities that may be obscure in the original context of virasoro blocks.

This paper is structured as follows. In section (\ref{sec:review}) we review some relevant facts about virasoro block in the large $c$  heavy-light limit, and gather some evidence for an underlying picture of phase transitions analogous to the Lee-Yang theory. In section (\ref{sec:AGT}) we invoke the AGT correspondence and cast the computation of the heavy-light virasoro block in terms of instanton partition functions. In section (\ref{sec:saddle}) we derive an effective theory at large instanton numbers, find a critical fugacity that matches the forbidden singularity, where instanton proliferation occurs and is described by a complex saddle-point solution. While not the focus of this paper, in section (\ref{sec:deconf}) we also study the phase on the other side of the critical fugacity, particularly the limiting behavior of the instanton gas that is related to the true singularity ($z=1$) of virasoro blocks. In section (\ref{sec:large_c}) we briefly comment on properties of the complex point $c=\infty$ as an essential singularity of the theory.  We conclude this paper with some discussion and outlooks in section (\ref{sec:discuss}).

\section{Semi-classical virasoro blocks}\label{sec:review}
We begin by reviewing in this section a few interesting facts regarding heavy-light virasoro blocks in the so-called semi-classical limit of $c\to \infty$, that are in connection with the phenomenon of ETH in 2d CFTs. Along the way, we will point to some ``phenomenological" evidences at the level of virasoro blocks, that nonetheless reveals both the existence and the nature of an underlying description involving phase transitions, whose explicit realizations we seek in the next section. 

The limit we are working with is termed semi-classical because it is believed that at the leading order in large $c$, the virasoro blocks exponentiates to the form: \footnote{From now on we shall focus only on a chiral factor of the virasoro block} 
\be\label{eq:semi_1}
\lim_{c\to \infty}\mathcal{V}^{h_i}_{h}(c, z) \sim e^{-\frac{c}{6} f(\epsilon_i,\epsilon_h,z)}
\ee
where $\epsilon_i = 6h_i/c$ are the rescaled external operator dimensions and $\epsilon_h = 6h/c$ is that of the internal dimension. In the holographic context, $1/c$ is identified with the Newton's constant $G_N$ in the bulk, and (\ref{eq:semi_1}) appears as the leading order semi-classical approximation of an quantum gravitational (bulk) computation, in which $\epsilon_i$ and $\epsilon_h$ are the effective bulk parameters.  

It is a daunting task to compute the virasoro blocks in closed-form expressions. A basic tool is the Zamolodchikov's recursive relation \cite{Zamolodchikov:1984eqp,Zamolodchikov:1987avt} that generates the coefficients $H_n$ in the following expansion: 
\bea\label{eq:recursive_1}
&&\mathcal{V}^{h_i}_h(c,z) = (16 q)^{h-\frac{c-1}{24}} z^{\frac{c-1}{24}}(1-z)^{\frac{c-1}{24}-h_2-h_3}\nonumber\\
&&\times \theta_3(q)^{\frac{c-1}{2}-4\sum_i h_i} H(c,h_i, h,q) \nonumber\\
&& q=e^{i\pi \tau},\;\tau = \frac{iK(1-z)}{K(z)},\;\theta_3(q)=\sum^{\infty}_{n=-\infty} q^{n^2}\nonumber\\
&&H(c,h_i,h,q)=\sum^{\infty}_{n=1} H_n\; q^n
\eea
The recursive method has been employed extensively in the studies, both analytically and numerically, in the study of semi-classical virasoro blocks \cite{Wang:2018,Chen:2017,Kusuki:2018wcv}.  

A more efficient method to directly extract the semi-classical behavior (\ref{eq:semi_1}) is through the so-called method of monodromy \cite{Fitzpatrick:2014,Hartman:2013,Harlow:2011ny}. The phenomenon of forbidden singularities are most conveniently accessed by this method. 

\subsection{Method of monodomy}
The method of monodromy is a systematic procedure to compute the semi-classical exponent $f(\epsilon_i,\epsilon_h,z)$ in (\ref{eq:semi_1}). It involves solving an auxiliary monodromy problem. To facilitate subsequent discussions, we provide a quick description of this procedure. For its derivation, see \cite{Fitzpatrick:2014,Harlow:2011ny}. The goal is to compute the virasoro block: 
\be
\lim_{c\to \infty} \mathcal{V}^{12,34}_h(c,z) = e^{-\frac{c}{6}f(\epsilon_i, \epsilon_h, z)}
\ee
One proceeds by solving the 2nd order ODE: 
\bea\label{eq:monodromy_ODE}
&&\Psi''(w)+U(w,z)\Psi(w) = \rho(w,z) \lambda \Psi(w)\nonumber\\
&& U(w,z) = \frac{\epsilon_1}{w^2}+\frac{\epsilon_2}{(w-z)^2}+\frac{\epsilon_3}{(w-1)^2}+\frac{\sum_i \epsilon_i -2 \epsilon_4}{w(1-w)}\\
&&\rho(w,z) = \frac{z(1-z)}{w(w-z)(1-w)}
\eea
where $\epsilon_i = 6h_i/c$ are the rescaled conformal dimensions of the external operators. The two linearly independent solutions $\Psi_{1,2}$ exhibit non-trivial monodromies around any of the regular singularities at $w=0,1,z,\infty$. To compute the virasoro block in the channel $\lbrace 1,2\to 3,4\rbrace$, we study the monodromy matrix $M$ around contours wrapping both singularities at $w=0$ and $w=z$. This monodromy matrix $M$ depends on the parameters of the ODE, in particular it depends on both the conformal ratio $z$ and the spectral parameter $\lambda$, i.e. $M = M(z,\lambda)$. The virasoro block is then computed by first solving the monodromy equation: 
\be\label{eq:monodromy_1}
\text{Tr}\;M(z,\lambda) =  -2 \cos{(\pi \Lambda_h)},\;\;\; h = \frac{c}{24}(1-\Lambda_h^2)  
\ee
A solution to (\ref{eq:monodromy_1}) amounts to finding $\lambda=\lambda(z)$ as a function of the conformal ratio $z$. This function is then related to the semi-classical virasoro block by: 
\be\label{eq:monodromy_2}
\partial_z f(\epsilon_i,\epsilon_h, z) = \lambda(z)
\ee
In particular, for the virasoro vacuum block $h=0$, the monodromy equation amounts to $M=\mathds{1}$, i.e. requiring both solutions to be periodic after traversing the contour. 

\subsection{Forbidden singularities}
The monodromy eqaution (\ref{eq:monodromy_1}) in its general form is a transcendental equation, making it difficult to solve in exact terms. Progress can be made in the heavy-light limit, i.e. $\epsilon_{1,2}=\epsilon_L=6h_L/c,\;\epsilon_{3,4}=6h_{H}/c$, by treating the light operator dimension $\epsilon_L$ as an expansion parameter, and solving the equation perturbatively in small $\epsilon_L$ \cite{Fitzpatrick:2014}. In particular, by self-consistently assuming that the accessory parameter $\lambda$ is also small, the vacuum monodromy equation $\text{Tr}\;M(z,\lambda) =2 $ can be expanded as a double series expansion in both small parameters $\lambda$ and $\epsilon_L$: 
\be\label{eq:monodromy_2}
\sum_{m,n\geq 1} M_{mn}(\epsilon_H,z) \;\lambda^{m} \epsilon_L^{n}=0
\ee
where the coefficients $M_{mn}(\epsilon_H,z)$ can be computed in principle systematically, though the computations become intractable beyond the leading orders. The fact that the expansion begins at $m+n=1$ means that treating the solution $\lambda$ as a small parameter for small $\epsilon_L$ is a consistent assumption.  At the leading order, we can truncate the equation (\ref{eq:monodromy_2}) at the order $m+n=1$: 
\be \label{eq:monodromy_3}
M_{10}(\epsilon_H,z)\lambda+M_{01}(\epsilon_H,z) \epsilon_L +... =0 
\ee
By plugging in the explicit expressions for $M_{01}$ and $M_{10}$ computed in \cite{Fitzpatrick:2014} one can solve for $\lambda(z)$ at the leading order in $\epsilon_L$: 
\bea\label{eq:forbidden_1}
\lambda(z) &\approx &  \frac{\left(1-i\alpha_H-(1-z)^{i\alpha_H}(1+i\alpha_H)\right)\epsilon_L}{(1-z)[(1-z)^{i\alpha_H}-1]}
\eea
where $\alpha_H = \sqrt{4\epsilon_H-1}$. In addition to $z=1$, this solution has additional singularities at where factor $(z_n-1)^{i\alpha_H}-1$ in the denominator vanishes: 
\be
z_n = 1-e^{-\frac{2\pi n}{\alpha_H}},n\in \mathds{Z}
\ee
They coincide with the forbidden singularities $z_n = 1-e^{-n\beta_H}$ by identifying $\beta_H = 2\pi/\alpha_H$, which is the BTZ black hole temperature matching the ADM energy of the state created by $O_H$. The result (\ref{eq:forbidden_1}) explicitly demonstrates the existence of forbidden singularities in the heavy-light limit of semi-classical virasoro block, and the locations agree with ETH. In fact, after integrating the accessory parameter (\ref{eq:forbidden_1}) over $z$ as prescribed, and transforming the result back onto the cylinder, the thermal expression in the high temperature limit is produced: 
\be 
\langle O_L(0) O_L(x)\rangle_{H} \approx \left[\frac{\beta_H}{\pi}\sin{\left(\frac{\pi x}{\beta_H}\right)}\right]^{-2h_L}
\ee

We can probe deeper about the nature of the forbidden singularities by considering further corrections from small $\epsilon_L$ effects  \cite{Wang:2018}. Re-summing over such corrections amounts to solving the transcendental equation (\ref{eq:monodromy_2}) exactly and is thus difficult, what is tractable is to re-summing a particular set of corrections that are important near $z=z_n$. In fact, the origin for the singularities of $\lambda(z)$ near $z_n$ can be identified as the vanishing of the monodromy equation coefficient $M_{10}(\epsilon_H,z)$ as $z\to z_n$. This makes the linear approximation of the monodromy equation (\ref{eq:monodromy_3}) degenerate. A natural thing to do is to include further corrections to the approximated monodromy equation, in this case a quadratic term in $\lambda$, so that near $z=z_n$ it becomes:
\be
-b_n \lambda^2 - \alpha_H(z-z_n) \lambda + 2\epsilon_L \alpha_H  \approx 0
\ee
giving rise to two branches of solutions:
\be
\lambda^{\pm}(z) \approx \frac{\alpha_H(z_n-z)\pm \sqrt{\alpha_H^2(z-z_n)^2+8b_n\epsilon_L \alpha_H}}{2b_n}
\ee
While the precise value of $b_n$ is not important, it is a positive real number. As a result, a more intricate structure of forbidden singularities emerges after re-summing a class of $\epsilon_L$ corrections that are important near the original locations $z=z_n$: instead of a pole at $z=z_n$, a pair of branch-points appears at: 
\be\label{eq:branch_pts}
\tilde{z}_n^{\pm} = z_n \pm i\sqrt{\frac{8 b_n \epsilon_L}{\alpha_H}}
\ee
while the original divergence at $z=z_n$ is regularized to $\lambda^{\pm}(z_n) \propto \sqrt{\epsilon_L} $. The nature of the forbidden singularities is drastically modified by such corrections: it changes from poles to branch-cuts, which are still forbidden. As solutions to the monodromy equation, $\lambda^{\pm}(z)$ correspond to the two sheets of a Riemann surface that are glued via the branch-cuts emanating from $z^\pm_n$. The multi-sheeted structure has a natural explanation from the monodromy equation: 
\be
\text{Tr}\;M(z,\lambda) = -2\cos{\left(\pi \Lambda_h\right)},\;\;h=\frac{c}{24}(1-\Lambda_h^2)
\ee
whose RHS is a periodic function of $\Lambda_h$. Due to this feature, infinitely many virasoro blocks with (un-physical) internal dimensions given by $h = -\frac{c}{6}n(n+1)$ solves the same vacuum monodromy equation \cite{Wang:2018,Fitzpatrick:2016mjq, Chen:2016cms, Benjamin:2023uib}. They provide the sheets that are connected by the branch-cuts. For more details about the branch-structure of the Riemann surface in relation to the un-physical blocks, see \cite{Wang:2018}.   

\subsection{Zero condensation and phase transition}
Based on the non-analyticity of $\lambda(z)=-\lim_{c\to \infty}\frac{6}{c}\partial_z \mathcal{V}^{t}_{vac}(c,z)/\mathcal{V}^t_{vac}(c,z)$ in the form of branch-cuts, emerged in the semi-classical limit of $c\to \infty$, one can make educated guesses on how it is resolved once $1/c$ corrections are fully recovered. A likely scenario is that at finite $c$, the branch-cut for $\lambda(z)$ becomes a trajectory of densely-packed poles. Such poles are admissible if they correspond to zeros of the virasoro block $\mathcal{V}^t_{vac}(c,z)$. In other words, we may expect at large but finite $c$ a trajectory of densely-packed zeros for $\mathcal{V}^t_{vac}(c,z)$. By numerically computing $\mathcal{V}^t_{vac}(c,z)$ at finite but large $c$ using the Zamolodchikov's recursive relation to sufficiently high orders, this was indeed observed \cite{Wang:2018}. As one increase $c$ in the heavy-light limit, the spacing between zeros vanish as $1/c$, while the distance between the real-axis and the zeros closest to it decrease as $1/\sqrt{c}$. In the $c\to \infty$ limit the  zeros coalesce into continuous curves that fall onto the real-axis, becoming branch-cuts for $\lambda(z)$. Via an analogy, this is very similar to how continuous phase-transitions arise in the Lee-Yang theory \cite{Yang:1952be,Lee:1952ig}. 

We elaborate on this analogy. We can view the conformal ratio $z$ as a (complex) tuning parameter (e.g. fugacity), that explores the phases of a system whose partition function is computed by $\mathcal{V}^t_{vac}(c,z)$, and $\lambda(z) \propto c^{-1}\partial_z \ln{\mathcal{V}^t_{vac}(x,z)}$ plays the role of an intensive thermodynamic observable (e.g. density). In the semi-classical limit $c\to \infty$, we can interpret the findings of the monodromy method as follows. The discontinuity of $\lambda(z)$ across the branch-cut $\gamma$ is analogous to a first-order phase transitions at complex $z$ between two saddle-point contributions to $\mathcal{V}^{t}_{vac}(c,z)$: 
\be 
\mathcal{V}^t_{vac}(c,z) \approx e^{-\frac{c}{6}f_1(z)} + e^{-\frac{c}{6} f_2(z)}
\ee
The distinct saddles-points realizes the multi-sheeted structure of $\lambda(z)$ found using the monodromy method, and so must be related to the un-physical blocks with $h_n = -cn(n+1)/6$, see \cite{Benjamin:2023uib,bissi20241cexpansion2dcfts} for related resurgence analysis in the $1/c$ expansion. The curve $\gamma$ of the first-order transitions is then described by the condition: 
\be
\text{Re} f_1(z) = \text{Re} f_2(z)
\ee
In the context of the WKB analysis, this is an anti-Stokes' curve. The condensation of zeros at large but finite $c$ may then show up along $\gamma$ as a natural phenomenon due to the destructive interference between both phases:  
\be
\mathcal{V}^t_{vac}(c,z) \sim e^{-\frac{6}{c} \text{Re} f_1(z)} \cos{\left[\frac{c}{6}\Big(\text{Im}f_1(z)-\text{Im}f_2(z)\Big)\right]}
\ee

The anti-Stokes curve in parametric form $\gamma=\lbrace z(t),\;t\in \mathds{R}^+\rbrace$ can be obtained from the solution $\lambda^{\pm}(z)$ of the monodromy problem by solving (numerically) the equation: 
\be\label{eq:branch_cuts}
\text{Re}\;\int^{z(t)}_{z(0)} dz' \left(\lambda^+(z') - \lambda^-(z')\right)=0
\ee
with the initial condition set by any of the branch-points $z(0) = z^\pm_n$, where $\lambda^{+}(z^{\pm}_n)=\lambda^-(z^{\pm}_n)$, thereby fixing a branch-cut emanating from $z^\pm_n$. From $\lambda^{\pm}(z)$ we can also write down the density of zeros on $\gamma$: 
\be
p(z) = \frac{\pi c}{6}\;\text{Im} \left(\lambda^{+}(z)-\lambda^{-}(z)\right) 
\ee

\begin{figure}[!htp]
\centering 
\includegraphics[scale=0.5]{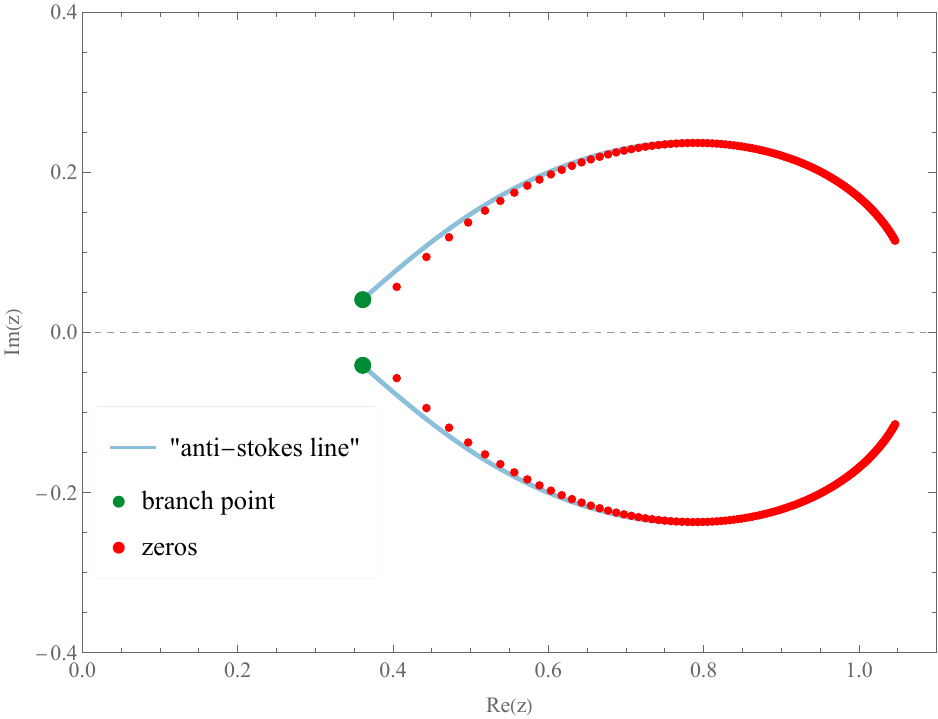}
\includegraphics[scale=0.5]{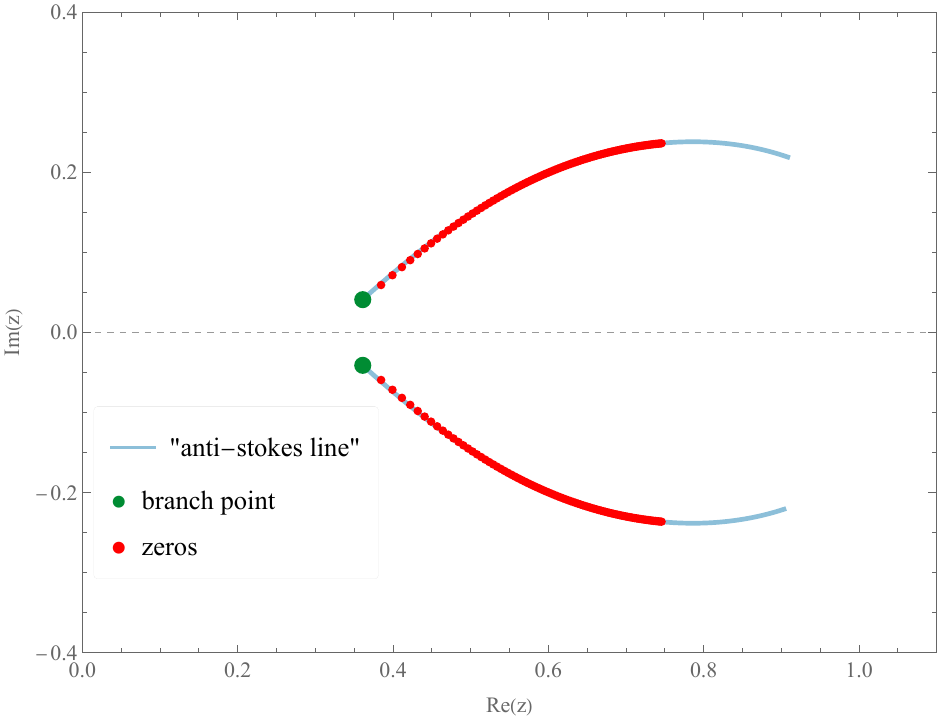}
\caption{Zeros (red dots) of the finite $c$ virasoro blocks, against the ``anti-stokes curve" $\gamma$ (blue), for fixed values of $\epsilon_L = 6*10^{-3},\;\;\epsilon_H = 60$. The values of $c$ are: $c=100$ (top) and  $c=400$ (bottom).}\label{fig:zeros}
\end{figure}

Eq (\ref{eq:branch_cuts}) provides a prediction for the locations of zeros at finite $c$ that should become more accurate as $c$ increases, if the above analogy is valid. In Figure (\ref{fig:zeros}) we plot the numerical solutions of $\gamma(t)$ obtained from (\ref{eq:branch_cuts}), against locations of zeros in the finite $c$ results, computed using Zamolodchikov's recursive relation to sufficiently high orders. The results indeed show good agreement that improves at large $c$, thus providing strong evidence for the validity of our analogy. An interesting observation is that $\gamma$ will cross the light-cone branch-cut from $z=1$ and extend into the Lorentzian sheets of $\mathcal{V}^t_{vac}(c,z)$. See \cite{Wang:2018} for a discussion on how this affects the real-time dynamics of holographic correlators.  

At the branch-points $z^{\pm}_n$, the phase transition changes from first-order to continuous. At finite $c$, $z^\pm_n$ are approximately the zeros that are closest to the real-axis. From the monodromy analysis (\ref{eq:branch_pts}), they appear in complex conjugate pairs whose distance to the real-axis is $\propto 1/\sqrt{c}$ in the heavy-light limit. Based on this, we expect that a continuous phase transition takes place at real and positive $z$, as the zeros of $\mathcal{V}^t_{vac}(c,z)$ at finite $c$ condense and falling onto the real-axis in the limit of $c\to \infty$, analogous to the Lee-Yang theory.

\begin{figure}[h]
\includegraphics[scale=0.55]{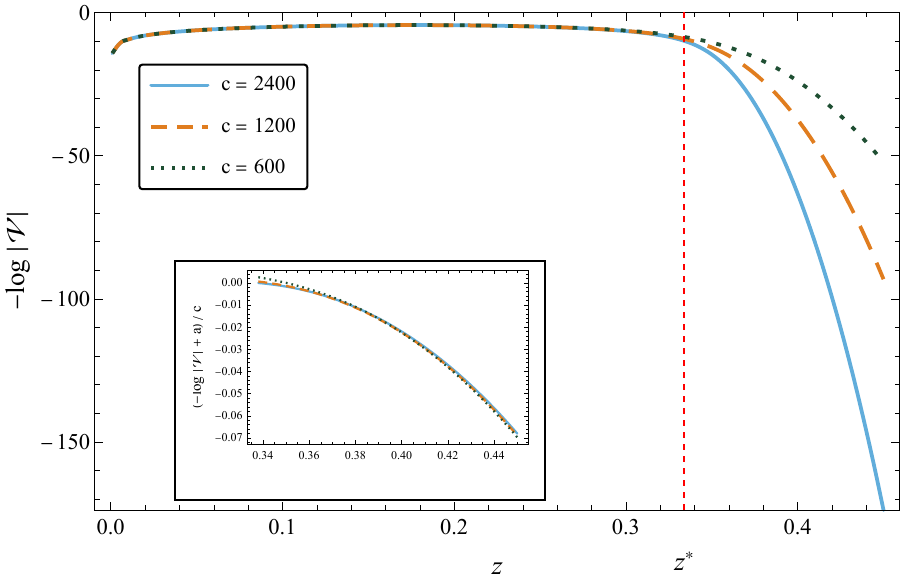}
\caption{Free energy at $\epsilon_H =60,\;\;h_L = 1$ as functions of $z$ for a different values of $c$. The scaling behaviors of $\mathcal{V}^t_{vac}(c,z)$ with respect to $c$ changes abruptly from $c$-independent to linear in $c$ (see the sub-figure, fitted with a constant shift $a\approx 10.46$) across $z^* \approx 0.334$.}\label{fig:deconfining}
\end{figure}

We can gain further insights by examining the behavior of the virasoro block $\mathcal{V}^t_{vac}(c,z)$ at finite but large $c$, as $z\in \mathds{R}^+$ increases from $z=0$ and cross the first forbidden singularity $z^*=z_1$. \footnote{For the finite $c$ block $\mathcal{V}^{t}_{vac}(c,z)$, only the first forbidden singularity at $z_1$ is accessible. The other forbidden singularities are associated with the related un-physical blocks $\mathcal{V}^t_{h_n}(c,z)$. This reflects the branching structure of the Riemman surface representing the monodromy solution $\lambda(z)$, see \cite{Wang:2018} for more details. } In Figure (\ref{fig:deconfining}) we plot the ``free energy" $\mathcal{F}(z)=-\ln{\mathcal{V}^t_{vac}(c,z)}$ in the heavy-light limit of large but finite $c$, computed also using Zamolodchikov's recursive relation. We see that the scaling behavior of $\mathcal{F}(z)$ changes abruptly across $z=z_1$: 
\be\label{eq:deconfine}
\mathcal{F}(z) \propto 
\begin{cases}
\mathcal{O}(1),\;\;\;\;0<z<z_1\\
\mathcal{O}(c),\;\;\;\;z_1<z<1
\end{cases}
\ee

This further supports via the analogy that there exists two distinct phases separated across $z^*$, in which the number of effective degrees of freedom is $\propto \mathcal{O}(1)$ in a ``confined" phase; and is liberated to $\propto \mathcal{O}(c)$ in a ``deconfined" phase. In the following sections, we make the analogy concrete.  

\section{Instanton partition function via AGT}\label{sec:AGT}
 It turns out there exists a natural framework, in which the virasoro blocks appears as actual partition functions of 4d gauge theories, it is the so-called AGT correspondence \cite{Akers:2020pmf,Nekrasov:2002qd,Nekrasov:2003rj}. It is an explicit realization of a broader class of 4d/2d correspondence, which relates 4d $\mathcal{N}=2$ SUSY gauge theories with 2d CFTs via a parent 6d $\mathcal{N}=(0,2)$ theory. For more details, see \cite{Tachikawa:2016kfc}.
 
 In this section, we apply the AGT correspondence to the particular context of heavy-light blocks. We begin by quoting some relevant aspects. Via the AGT correspondence, a virasoro conformal block $\mathcal{V}$ is found to be equivalent to the partition function of certain 4d $\mathcal{N}=2$ SUSY $\text{SU}(2)$ gauge theory:
\begin{align}\label{eq:AGT}
&{\cal V}\Bigl[
{\fontsize{8pt}{8pt}\selectfont
\begin{matrix}
2&3\\
1&4
\end{matrix}
}
;\Delta,z
\Bigr]={\cal Z}_{\text{4d}}(z),\quad {\cal Z}_{\text{4d}}(z)=\frac{\mathcal{Z}^{\text{U}(2)}_{\text{inst}}(z)}{\mathcal{Z}^{\text{U(1)}}(z)}
\end{align}

Let us unpack the equivalence relation (\ref{eq:AGT}). The LHS is the virasoro block with external dimensions $\Delta_i$ and internal dimension $\Delta$ in the OPE channel $(\Delta_1,\Delta_2)\to \Delta \to  (\Delta_3,\Delta_4)$, evaluated at conformal ratio $z$ at central charge $c$. The RHS is the 4d partition function of $\mathcal{N}=2$ SUSY gauge theory with gauge group $\text{SU}(2)$ and four hyper multiplets of masses $\mu_{i=1,...,4}$ at the point $(a,-a)$ of the Coulomb-branch moduli space, evaluated on the $\Omega$-deformed background with deformation parameters $\lbrace \epsilon_1,\epsilon_2\rbrace$.  The gauge theory parameters are completely determined by the CFT parameters in a way that is summarized in Table (\ref{tab:AGT_parameters}). 
\begin{table}[htbp]
  \centering
  \label{tab:centered}
  \begin{tabular}{|c|c|}   % both columns centered, with vertical lines
    \hline
    \textbf{2d virasoro block} & \textbf{4d partition function} \\
    \hline
    conformal ratio $z$    &  gauge coupling $q = \frac{\theta}{2\pi}+\frac{4\pi i}{g^2}$  \\
    \hline
    external dimensions $\Delta_{1,...,4}$      & hyper multiplet masses $\mu_{i=1,...,4}$ \\
    \hline
    internal dimension $\Delta$   & Coulomb-branch moduli $a$ \\
    \hline
    central charge $c$     & $\Omega$-deformation parameters $\lbrace \epsilon_1,\epsilon_2\rbrace$ \\
    \hline
  \end{tabular}
  \caption{Relations between CFT and gauge theory parameters in AGT correspondence}\label{tab:AGT_parameters}
\end{table}
The explicit conversion formulae between gauge theory and CFT parameters are given by: (\ref{eq:AGT_conversion}):
\begin{align}\label{eq:AGT_conversion}
&z=e^{2\pi i q},\;\;c=1+6Q^2,\quad Q=
\sqrt{\frac{\epsilon_1}{\epsilon_2}}
+\sqrt{\frac{\epsilon_2}{\epsilon_1}}\nonumber\\
&\quad \Delta_i=\alpha_i(Q-\alpha_i), \quad\Delta = \alpha(Q-\alpha) \nonumber \\
&\frac{\mu_1}{\sqrt{\epsilon_1\epsilon_2}}=\frac{Q}{2}-\alpha_1+\alpha_2,\quad
\frac{\mu_2}{\sqrt{\epsilon_1\epsilon_2}}=-\frac{Q}{2}+\alpha_1+\alpha_2,\quad\nonumber\\
&\frac{\mu_3}{\sqrt{\epsilon_1\epsilon_2}}=\frac{Q}{2}-\alpha_3+\alpha_4,\frac{\mu_4}{\sqrt{\epsilon_1\epsilon_2}}=-\frac{Q}{2}+\alpha_3+\alpha_4\nonumber\\
&\quad \alpha=\frac{Q}{2}+\frac{a}{\sqrt{\epsilon_1\epsilon_2}},\quad (a_1,a_2)=(a,-a)
\end{align}
The $\text{SU}(2)$ gauge theory partition function On the RHS of (\ref{eq:AGT}) is given by the ratio between a $\text{U}(2)$ and a $\text{U}(1)$ factors. The $\text{U}(1)$ factor $\mathcal{Z}^{\text{U}(1)}
(z)$ is simply given by  $(1-z)^{f_{\text{U}(1)}}$ with exponent:
\bea\label{eq:U1_factor}
f_{\text{U}(1)}&=& 2Q^2 + Q(\alpha_1 - 3 \alpha_2 + \alpha_3 - 3 \alpha_4) \nonumber\\
&-& \alpha_1^2 + \alpha_2^2 - \alpha_3^2 + \alpha_4^2 + 2 \alpha_2 \alpha_4
\eea
The interesting features of the virasoro block that we aim to investigate, e.g. forbidden singularities, are entirely encoded in the $Z_{\text{inst}}^{\text{U}(2)}$ factor. We focus on it from now on. Being related to the virasoro block, it also admits a series expansion in $z=e^{2\pi i \tau}$:
\be
\mathcal{Z}_{\text{inst}}^{\text{U}(2)}(z)=
\sum_{\nu=0}^\infty z^\nu\times \mathcal{Z}_{\text{inst},\nu}^{\text{U}(2)}
\ee
In terms of the SUSY gauge theory the expansion in $z=e^{2\pi i q}$ is essentially the fugacity expansion of the instanton gas, where $\nu$ is the instanton number and $z$ is the effective fugacity. Using the technique of SUSY localization, computing the coefficient $\mathcal{Z}^{\text{U(2)}}_{\text{inst},\nu}$ can be reduced onto fixed-points of the $\nu$-instanton moduli-space, which is rendered smooth by the non-commutative effects introduced through the $\Omega$-deformation. In this case, each fixed-point configuration is a micro-state, and is represented by a pair of Young tableaux $Y^{1,2}$ with $|Y^1|+|Y^2|=\nu$, where $|Y|$ denotes the total size, i.e. number of boxes, of $Y$. The corresponding micro-state contribution to $\mathcal{Z}^{\text{U(2)}}_{\text{inst},\nu}$ can be factorized into the vector-multiplet part and the hyper-multiplet part, giving rise to the following prescription:
\bea
\mathcal{Z}_{\text{inst},\nu}^{\text{U}(2)}&=& \sum_{|Y^1|+|Y^2| =\nu } e^{-\mathcal{I}_{\text{inst}}(Y^1,Y^2)}\nonumber\\
e^{-\mathcal{I}_{\text{inst}}(Y^1,Y^2)}&=&\frac{\prod^{4}_{i=1} \mathcal{Z}_{\text{fund}}(Y^{1,2},\mu_i)}{\mathcal{Z}_{\text{vec}}(Y^{1,2})}
\eea
Each part can be explicitly written as a product over  factors associated with the boxes $\Box\in Y^{1,2}$:
\bea\label{eq:Nekrasov}
&&\mathcal{Z}_{\text{vec}}(Y^{1,2})=\prod_{\alpha,\beta=1}^2\Bigg[
\prod_{\Box\in Y^\alpha}\bigg(a_\alpha-a_\beta-\ell_{Y^\beta}(\Box)\epsilon_1\nonumber\\
&+&(a_{Y^\alpha}(\Box)+1)\epsilon_2\bigg)
\prod_{\Box\in Y^\beta}\bigg(a_\alpha-a_\beta+(\ell_{Y^\alpha}(\Box)+1)\epsilon_1\nonumber\\
&-& a_{Y^\beta}(\Box)\epsilon_2\bigg)
\Bigg]\nonumber \\
&& \mathcal{Z}_{\text{fund}}(Y^{1,2},\mu)=\prod_{\alpha=1}^2\prod_{\Box\in Y^\alpha}(a_\alpha+\epsilon_1 i+\epsilon_2 j-\mu)
\eea
We explain (\ref{eq:Nekrasov}) by clarifying the following notations and rules:
\begin{itemize}
\item A Young diagram is specified by a non-increasing sequence of positive integers $Y=(Y_1\ge Y_2\ge \cdots\ge Y_d>0)$. They represent the lengths of (i.e. number of boxes in) horizontal rows of $Y$, in which there are $d$ such rows. In this notation, the total size of $Y$ is given by: $|Y|=\sum^{d}_{i=1} Y_i$.

\item Each box $\Box = (i,j) \in Y$ is specified by its row index $i$ and column index $j$, i.e. $\Box=(i,j)\in Y$ is located at the $j$-th box in the $i$-th row of $Y$, counted from the top-left corner of $Y$. 

\item For each box $s=(i,j)\in Y$, we define its arm-length $a_Y(s)$ and leg-length $\ell_Y(s)$ that appear in (\ref{eq:Nekrasov}) as follows:
\be\label{eq:YD_armlegs}
a_Y(s)=
Y_i-j,\;\;
\ell_Y(s)= X_j-i
\ee
where $X={}^t Y$ denotes the transpose of $Y$, i.e. obtained from $Y$ by switching its rows and columns. In terms of the original $Y$, $X_j$ denotes the column length of the $j$-th column. For a diagrammatic illustration, see Figure (\ref{fig:YD}):
\begin{figure}[!htp]
\centering
\includegraphics[scale=0.95]{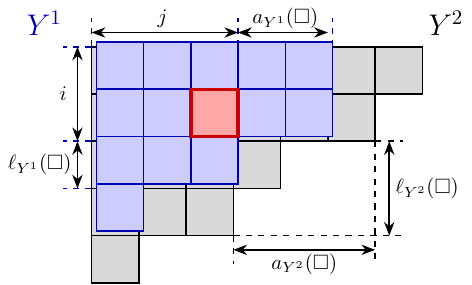}
\caption{An illustrative example of determining $\lbrace i,j,\ell_{Y},a_{Y}\rbrace$ for $\Box \in Y^1$ (red), evaluated in $Y^1$ (blue) and $Y^2$ (grey).}\label{fig:YD}
\end{figure}

\item Notice that when evaluating $\mathcal{Z}_{\text{vec}}$ in (\ref{eq:Nekrasov}), we need to specify the ``off-diagonal" cases for : 
\be 
a_{Y^\alpha}(\Box),\;\ell_{Y^\alpha}(\Box),\;\;\Box\in Y^\beta,\;\alpha \neq \beta
\ee
This is defined also according to (\ref{eq:YD_armlegs}). In doing this, the off-diagonal arm/leg-length may take negative integer values if $\Box =(i,j)$ is outside $Y^\alpha$. 

\item To accommodate the definition (\ref{eq:YD_armlegs}) for the leg-length $\ell_Y(\Box)$ when $\Box=(i,j)$ is to the right of $Y$, i.e. when $j>Y_1$, we also define $X_j=0$ for $j>Y_1$. 
\end{itemize}

To summarize, the virasoro block is described through the AGT correspondence by the grand canonical partition function of instanton gas, where $z=e^{2\pi i \tau}$ plays the role of fugacity in particle number $\nu$. The micro-states of the instanton gas are specified by a pair of Young tableaux $Y^{1,2}$. In more precise terms, it is analogous to a ``solution" consisting of two species of particles in chemical equilibrium, i.e. sharing the same fugacity $z$. To make connection with the monodromy analysis of virasoro block in the previous section, we notice that the accessory parameter $\lambda(z) = 6 c^{-1}\partial_z \mathcal{F}(z)$ is related to the average density (in the sense of ``per central charge")  of the instanton gas at fugacity $z$. Here we have treated the central charge $c$ as the effective volume for the instanton gas emergent from AGT.    

\section{Instanton proliferation: phase transition and forbidden singularity}\label{sec:saddle}

Solving properties of the instanton gas exactly through (\ref{eq:Nekrasov}) is equivalent to computing the virasoro block exactly, and is beyond the scope of this work. We are only interested in its properties related to the emergence of forbidden singularities of $\mathcal{V}^t_{vac}(c,z) \propto \mathcal{Z}^{\text{U(2)}}_{\text{inst}}(z)$ in the heavy-light limit. From a mathematical point of view, a singularity located at $|z|=r$ implies that the low fugacity expansion: 
\be\label{eq:instanton_expansion}
\mathcal{V}^t_{vac}(c,z) \propto \mathcal{Z}^{\text{U(2)}}_{\text{inst}}(z)  = \sum^\infty_{\nu=0} z^{\nu} \times \mathcal{Z}^{\text{U(2)}}_{\text{inst},\nu} 
\ee
has a radius of convergence $r$ determined by: 
\be\label{eq:radius} 
r = \lim_{\nu\to \infty} \left(\frac{\mathcal{Z}^{\text{U(2)}}_{\text{inst},\nu}}{\mathcal{Z}^{\text{U(2)}}_{\text{inst},\nu+1}} \right) 
\ee
The value of $r$ therefore dictates the critical fugacity at which the low fugacity expansion breaks down and gives rise to a singularity of the grand canonical partition function. To declutter notations, from now on we write $\mathcal{Z}^{\text{U(2)}}_{\text{inst},\nu}$ simply as $\mathcal{Z}_{\nu}$. In practice, the radius of convergence (\ref{eq:radius}) can be extracted from the dominant growth of the asymptotic coefficients: \be\label{eq:f_scaling}
r  = \lim_{\nu \to \infty} \left(\mathcal{Z}_{\nu}\right)^{-1/\nu}
\ee
As the fugacity approaches the critical value $z\to r$, the average instanton number $\langle \nu \rangle = z\partial_z \ln{\mathcal{Z}^{\text{SU(2)}}_{\text{inst}}(z)}$ diverges. Physically, this is a signal for the proliferation of instantons, and marks a potential instability of the gas.  

In parallel with the contrast between the forbidden singularity and true OPE singularity of the virasoro block, here one also encounters two notions of convergence radius: one emerges only in the $c \to \infty$ limit and corresponds to the forbidden singularity $z^*$; the other is valid at finite $c$ and corresponds to the true OPE singularity closest to $z=0$, in this case $z=1$. They lead to distinct predictions regarding $r$: \footnote{In more precise computations, $r=1$ can be realized by a sub-linear growth in the exponent, e.g. $\mathcal{Z}_\nu \sim e^{A \sqrt{\nu}}$ or $\mathcal{Z}_{\nu} \sim \nu^\gamma$, see \cite{Kusuki:2018wcv} for investigations of such behaviors in the $q$-expansion of virasoro block at finite $c$.}
\be\label{eq:kappa_distinct}
r = \begin{cases}
\;\;z^*\;\;\;\;c= \infty\\
\;\;1\;\;\;\;\;c <\infty
\end{cases}
\ee

What reconciles between the two predictions in (\ref{eq:kappa_distinct}) is a refined notion of large $\nu$ in relation to large but finite $c$. We remind that $\mathcal{Z}_\nu$ are $c$-dependent through the parameters $\epsilon_{1,2}$ in (\ref{eq:Nekrasov}). The distinct natures of $r$ in relation to large $c$ in (\ref{eq:kappa_distinct}) correspond to two distinct orders of limits: 
\bea\label{eq:order_limits}
&&\lim_{\nu \to \infty} \left[\lim_{c\to \infty} \left(\frac{\mathcal{Z}_\nu}{\mathcal{Z}_{\nu+1}}\right)\right] =z^* \nonumber\\
&&\lim_{c \to \infty} \left[\lim_{\nu\to \infty} \left(\frac{\mathcal{Z}_\nu}{\mathcal{Z}_{\nu+1}}\right)\right]= 1
\eea
In parametric terms, they are controlled by the regimes $1\ll \nu \ll c$ and $1\ll c\ll \nu$ respectively. In this sense, the phenomenon of forbidden singularity comes from the fact that the limits $c\to \infty$ and $\nu \to \infty$ do not commute. To probe the first order of limits in (\ref{eq:order_limits}), which reveals the forbidden singularity, we need to first send $c\to \infty$ for the parameters in (\ref{eq:Nekrasov}) and derive a limiting effective theory, then study the large instanton number $\nu$ limit of the effective theory. \footnote{While generic semi-classical blocks lead to the so-called Nekrasov-Shatashvili (NS) limit of the instanton partition functions \cite{Nekrasov:2009rc}, they are usually defined with $\lbrace \Delta_i, \Delta\rbrace \sim c$, and are controlled by regimes distinct from ours.} 

\subsection{Effective theory in the heavy-light limit}

In physical terms, the instanton gas described by (\ref{eq:Nekrasov}) is a complicated interacting system between the Young tableaux  $(Y^1,Y^2)$, each of which also features complicated self-interactions. The general structure of interactions can be broken down as follows. The self-interaction of $Y^\alpha$ is captured by the factors in $\mathcal{Z}_{\text{fund}}$, as well as the diagonal factors of $\alpha =\beta$ in $\mathcal{Z}_{\text{vec}}$. The interaction between $(Y^1, Y^2)$ is first encoded implicitly in the constraint $|Y^1|+|Y^1|=\nu$, and then explicitly in the off-diagonal factors of $\alpha \neq \beta$ in $\mathcal{Z}_{\text{vec}}$. 

It is clearly difficult to treat the interactions as they appear in the exact formula (\ref{eq:Nekrasov}). We now take its heavy-light limit of sending $c\to \infty$ while fixing: \footnote{The formula (\ref{eq:Nekrasov}) for the vacuum block at exactly $\Delta =0$ contains zero-factors $\propto \frac{Q}{2}\sqrt{\epsilon_1 \epsilon_2} - \epsilon_1-\epsilon_2 = 0$ that cancel out between the denominator and the numerator. To simplify things, in what follows we assume a generic non-zero value of $\Delta$ that is $\mathcal{O}(1)$ in the $c\to \infty$ limit. This modification is negligible for the discussions related to forbidden singularities.}
\be\label{eq:heavy_light}
\Delta_{1,2}=h_L \sim \mathcal{O}(1),\;\;\;\Delta_{3,4} = h_H = \frac{c}{6} \epsilon_H
\ee
Via the conversion formulae (\ref{eq:AGT_conversion}) of the AGT correspondence, the heavy-light limit (\ref{eq:heavy_light}) translates into the gauge theory parameters as the following large $c$ expansions:
\bea\label{eq:AGT_large_c}
&&\epsilon_1 = c, \;\;\epsilon_2 \to 6+78c^{-1}+\mathcal{O}(1/c^2)\nonumber\\
&&a=\frac{c}{2}+(3-6\Delta)+\mathcal{O}(c^{-1},\Delta^2)\nonumber\\
&&\mu_{1,3}=\frac{c}{2}+3+\mathcal{O}(1/c),\;\;
\mu_2 = \frac{3c}{2}+(9-12h_L)+\mathcal{O}(1/c)\nonumber\\
&&\mu_{4}=\left(\frac{1}{2}+i\alpha_H\right)c+\mathcal{O}(1)
\eea
In this limit the $U(1)$ factor has an $\mathcal{O}(1)$ exponent:
\be
f_{U(1)} = h_L\left(1-i\alpha_H\right)+ \mathcal{O}(c^{-1})
\ee
and plays no significant role in any aspect. It will not appear in our discussion from now on.

To derive the effective action of the instanton gas in the heavy-light limit, 
we plug (\ref{eq:AGT_large_c}) into the micro-state contribution (\ref{eq:Nekrasov}), keeping only the leading-order term at large $c$ in each term, and extracting:  
\be\label{eq:eff_def}
\lim_{c\to \infty} \mathcal{I}_{\text{inst}}(Y^1,Y^2)= \mathcal{I}_{0}(Y^1,Y^2) 
\ee

To proceed, we notice that in taking (\ref{eq:AGT_large_c}), a typical contribution to $\mathcal{I}_{\text{inst}}(Y^1,Y^2)$, associated with a box $\Box\in Y^{1,2}$, takes the general form: 
\be\label{eq:typical_1}
\pm\ln{\left( A c+ V(\Box) c + H(\Box) \right)}
\ee
where $A$ is some constant, $V(\Box)$ is an integer that depends on the vertical position of $\Box$, and $H(\Box)$ is an integer that depends on the horizontal position of $\Box$. For each $\Box\in Y^{1,2}$, there are in total 16 such terms from (\ref{eq:Nekrasov}), in which 8 terms from $\mathcal{Z}_{\text{vec}}$ contribute with the $(+)$ sign, and 8 terms from $\mathcal{Z}_{\text{fund}}$ contribute with the $(-)$ sign. For a generic $\Box\in Y^{1,2}$ with $|Y^1|+|Y^2|\ll c$, we have that $V(\Box), H(\Box) \ll c$, and the large $c$-expansion of (\ref{eq:typical_1}) is : 
\be
\pm \ln{c} \pm \ln{\left( A+ V(\Box)\right)} + \mathcal{O}(1/c)
\ee
for all 16 terms associated with $\Box$. Summing over all these contributions, the $\ln{c}$ terms cancel between those from $\mathcal{Z}$, and the leading-order term only depends on the vertical position of $\Box$. Had this been true for all $\Box \in Y^{1,2}$, we would have reached a very simple result for $\mathcal{I}_{0}$ that is $c$-independent and given by a sum over contributions from the columns of $Y^{1,2}$ that are decoupled. 

There are however special terms in which $A+V(\Box)=0$ when written as (\ref{eq:typical_1}). They give rise to contributions of the form $ \pm \ln{H(\Box)}$ that may not only introduce dependence on the horizontal position of $\Box$ -- thereby violating the decoupling between columns, but also pollute the cancellation of the $\ln{c}$ terms since they do not have one, causing the result to be $c$-dependent. 

Based on the specific assignment of the parameter values in (\ref{eq:AGT_large_c}), we identify the following cases when this can happen. For $\Box=(i,j)\in Y^1$, its contribution to $\mathcal{I}_{0}(Y^1,Y^2)$ contains a special term when:
\begin{itemize}
\item $i=1,\;\;\;\;\;\;\;\;\;\;\;\;\;\;\;(-1)$
\item $\ell_{Y^1}(\Box)=0,\;\;\;\;\;(+1)$
\item $\ell_{Y^2}(\Box)=0,\;\;\;\;\;(+1)$
\item $\ell_{Y^2}(\Box)=1,\;\;\;\;\;(+1)$
\end{itemize}
For  $\Box=(i,j)\in Y^2$, its contribution to $\mathcal{I}_{0}(Y^1,Y^2)$ contains a special term when:
\begin{itemize}
\item $i=1,\;\;\;\;\;\;\;\;\;\;\;\;\;\;\;(-2)$
\item $i=2,\;\;\;\;\;\;\;\;\;\;\;\;\;\;\;(-1)$
\item $\ell_{Y^2}(\Box)=0,\;\;\;\;\;(+1)$
\item $\ell_{Y^1}(\Box)=-1,\;\;(+1)$
\item $\ell_{Y^1}(\Box)=-2,\;\;(+1)$
\end{itemize}
where we have indicated the coefficient of the $\ln{c}$ term it ``removes" from a perfect cancellation. Notice that the special terms are associated with boxes of a Young diagram that are near the boundaries of either itself or of the other one ``immersed" in itself. 

\begin{figure}[!htp]
\centering
\includegraphics[scale=0.4]{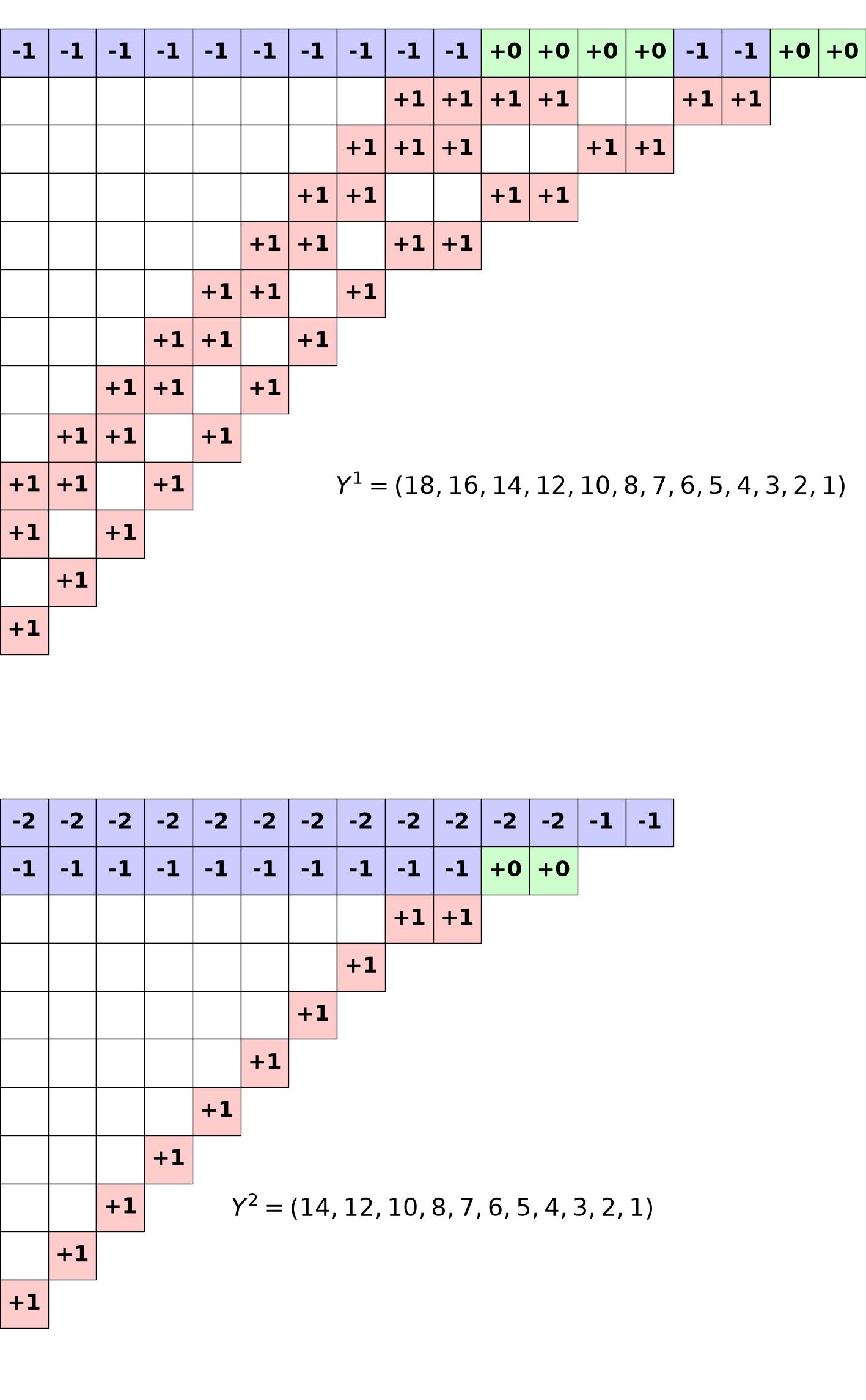}
\caption{An example of how the missing $\ln{c}$ terms cancel among contributions from the colored boxes, which contain the special terms. The total coefficient of the missing $\ln{c}$ for each colored box -- one could contain multiple special terms -- is shown in the box. They are colored differently depending whether it is net positive (red), net negative (purple), or net neutral (green). One can verify in this example that the total positive coefficients add up to $48$, while the total negative coefficients add up to $-48$, thereby managing a perfect cancellation. }\label{fig:cancellation}
\end{figure}

Remarkably, we can perform a counting analysis to show that the coefficients of the ``missing" $\ln{c}$ terms also cancel among the special terms. As a result, the total result for $\mathcal{I}_0(Y^1,Y^2)$ remains $c$-independent! In other words, (\ref{eq:eff_def}) has a well-defined $c\to \infty$ limit. We leave the detailed analysis in the Appendix (\ref{app:cancellation}). Here we illustrate the cancellation explicitly for a typical configuration $Y^{1,2}$, see Figure (\ref{fig:cancellation}). We emphasize that the conditions for these special terms depend crucially on the heavy-light arrangement of the external operators. For example, in the all-heavy limit, the conditions for the special terms are different, and the cancellation of $\ln{c}$ terms is no longer valid. 

The remaining computation is tedious but straightforward. In the end, the exact expression of $\mathcal{I}_0(Y^1,Y^2)$, valid for any $Y^{1,2}$ satisfying $|Y^1|+|Y^2|=\nu \ll c$, can be worked out in terms of two sequences of the row-lengths $Y^{1,2}_X$. More specifically, $\mathcal{I}_0(Y^1,Y^2)$ describes two chains with integer degrees of freedom $Y^{1,2}_X$ defined on sites labeled by the row-index $X$. They dynamics consist of three parts: the self-energies, the interaction, and the phase index: 
\bea\label{eq:effective_action}
\mathcal{I}_0(Y^{1,2})=\sum^{2}_{\alpha=1}\mathcal{I}_\alpha(Y^\alpha)+\mathcal{I}_{\text{int}}(Y^{1,2})+i\pi\Theta
\eea
The exact expressions of these terms are not important for our subsequent analysis -- they will simplify significantly later when we consider the large instanton number $\nu$ limit. For completeness, we reproduce them here. The two self-energy terms are given by:
\bea
&&\mathcal{I}_{1}(Y^1)= \sum_X N^1_X \ln{\left[\frac{\Gamma(X+2)\Gamma(1-i\alpha_H)}{\Gamma(X+1-i\alpha_H)}\right]}\nonumber\\
&+& \ln{\left[\frac{\Gamma(2h_L-\Delta)}{\Gamma\left(Y^1_1+2h_L-\Delta\right)}\right]}+\sum_{X} \ln{\Gamma(N^1_X+1)}
\eea
for $\mathcal{I}_1(Y^2)$ and: 
\bea 
&&\mathcal{I}_{2}(Y^2) = \sum_X N^2_X \ln{\left[\frac{\Gamma(X+1)\Gamma(-i\alpha_H)}{\Gamma(X-i\alpha_H)}\right]}\nonumber\\
&+&2\ln{\left[\frac{\Gamma(\Delta)}{\Gamma\left(Y^2_1+\Delta\right)}\right]}+\ln{\left[\frac{\Gamma(2h_L+\Delta-1)}{\Gamma\left(Y^2_2+2h_L+\Delta-1\right)}\right]}\nonumber\\
&+& \sum_{X} \ln{\Gamma(N^2_X+1)} 
\eea
for $\mathcal{I}_2(Y^2)$, from which we can identify terms that look like self-interactions as well as external potentials. The interaction term $\mathcal{I}_{\text{int}}(Y^{1,2})$ is given by: 
\bea
&&\mathcal{I}_{\text{int}}(Y^{1,2}) = \sum_{X} \ln\Bigg[\frac{\Gamma\left((Y^1_X-Y^2_{X+1})^++1-2\Delta\right)}{\Gamma\left((Y^1_X-Y^{2}_{X})^++1-2\Delta\right)} \nonumber\\
&& \frac{\Gamma\left((Y^1_X-Y^2_{X+2})^++2-2\Delta\right)\Gamma\left((Y^2_{X}-Y^1_{X})^++2\Delta\right)}{\Gamma\left((Y^1_X-Y^{2}_{X+1})^++2-2\Delta\right)\Gamma\left((Y^2_{X}-Y^{1}_{X-1})^++2\Delta\right)}\nonumber\\
&&\frac{\Gamma\left((Y^2_X-Y^1_{X-1})^+-1+2\Delta\right)}{\Gamma\left((Y^2_X-Y^{1}_{X-2})^+-1+2\Delta\right)}\Bigg]\nonumber
\eea
Finally the phase index $\Theta$ is given by:
\bea 
&&\Theta =Y^1_1-Y^2_2+(Y^2_1-\max\{Y^1_1,Y^2_2\})^+\nonumber\\
&+& \sum_{X}(\min\{Y^1_X,Y^2_{X+1}\}-\max\{Y^1_{X+1},Y^2_{X+2}\})^+
\eea
It counts the number of $(-1)$ factors, and is essentially determined by how the lower-right boundaries of $Y^1$ and $Y^2$ cross one another, which we call their crossing-pattern. In writing these expressions we have defined:
\bea
N^{1,2}_X = Y^{1,2}_{X}-Y^{1,2}_{X+1},\;(A)^+=\begin{cases}
A,\;\;\;A\geq 0\\
0,\;\;\;A<0
\end{cases}
\eea
The details of the derivation can be referred to in the Appendix (\ref{app:effective_action}). 

 The fact that the effective action is manifestly independent of $c$ is compatible with the low fugacity phase of (\ref{eq:deconfine}), in which the free energy $\mathcal{F}(z)$ is $c$-independent. We expect that in this phase, the free energy can be reliably approximated in the fugacity expansion using coefficients $\mathcal{Z}_\nu$ that are computed from the $c$-independent effective action $\mathcal{I}_0(Y^1,Y^2)$: 
\be\label{eq:eff_fugacity_exp} 
\lim_{c\to \infty} e^{-\mathcal{F}(z)}= \sum_\nu z^{\nu} \sum_{|Y^1|+|Y^2|=\nu} e^{-\mathcal{I}_0(Y^1,Y^2)}
\ee
This is valid as long as it converges, i.e. when $|z|$ is smaller than its radius of convergence. Outside the radius of convergence, the effective action (\ref{eq:effective_action}) breaks down. From a finite $c$ perspective, this could happen because the original fugacity expansion should be dominated by instanton gas configurations with $\nu\gtrsim c$, thereby invalidating the assumption of  $\nu \ll c$ for deriving (\ref{eq:effective_action}). Within the effective action (\ref{eq:effective_action}), entering the regime of $\nu \gtrsim c$ could appear as an instability of instanton proliferation towards $\nu \to \infty$. The goal now is to determine the critical fugacity, or the radius of convergence, for the effective fugacity expansion using (\ref{eq:effective_action}). If it reproduces the forbidden singularity $z^*$, then we have successfully found its underlying physical origin in terms of the instanton gas.  

\subsection{Statistical theory at large $\nu$}
As was outlined in (\ref{eq:radius}), to determine the radius of convergence for the effective fugacity expansion (\ref{eq:eff_fugacity_exp}) we need to compute the large $\nu$ limit of: \be \label{eq:coeff_large_c}
\bar{\mathcal{Z}}_\nu \equiv \lim_{c\to \infty}\mathcal{Z}_\nu = \sum_{|Y^1|+|Y^2|=\nu} e^{-\mathcal{I}_0(Y^1,Y^2)}
\ee 
Since $c$ has effectively disappeared in $\bar{\mathcal{Z}}_\nu$, the restriction of $\nu \ll c$ is always implicitly maintained when computing (\ref{eq:coeff_large_c}) at large $\nu$. The number of Young tableaux configurations satisfying $|Y^1|+|Y^2|=\nu$ grows very quickly, and the only viable approach to study the sum over $Y^{1,2}$ in (\ref{eq:coeff_large_c}) at large $\nu$ is to work out a statistical theory regarding the dominant contributions.

\begin{figure}[!htp]
    \centering
    \includegraphics[scale=0.21]{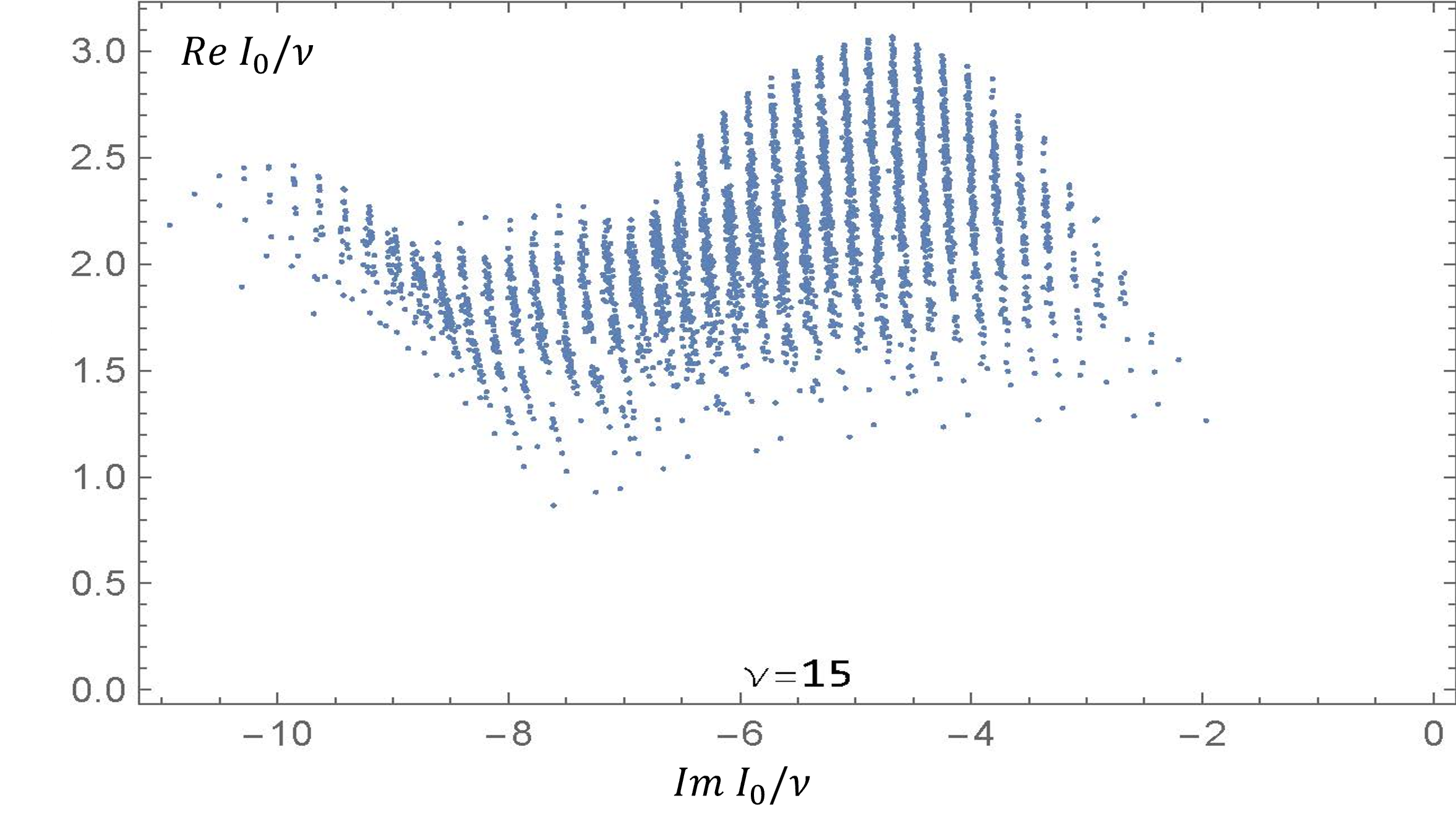}
    \includegraphics[scale=0.21]{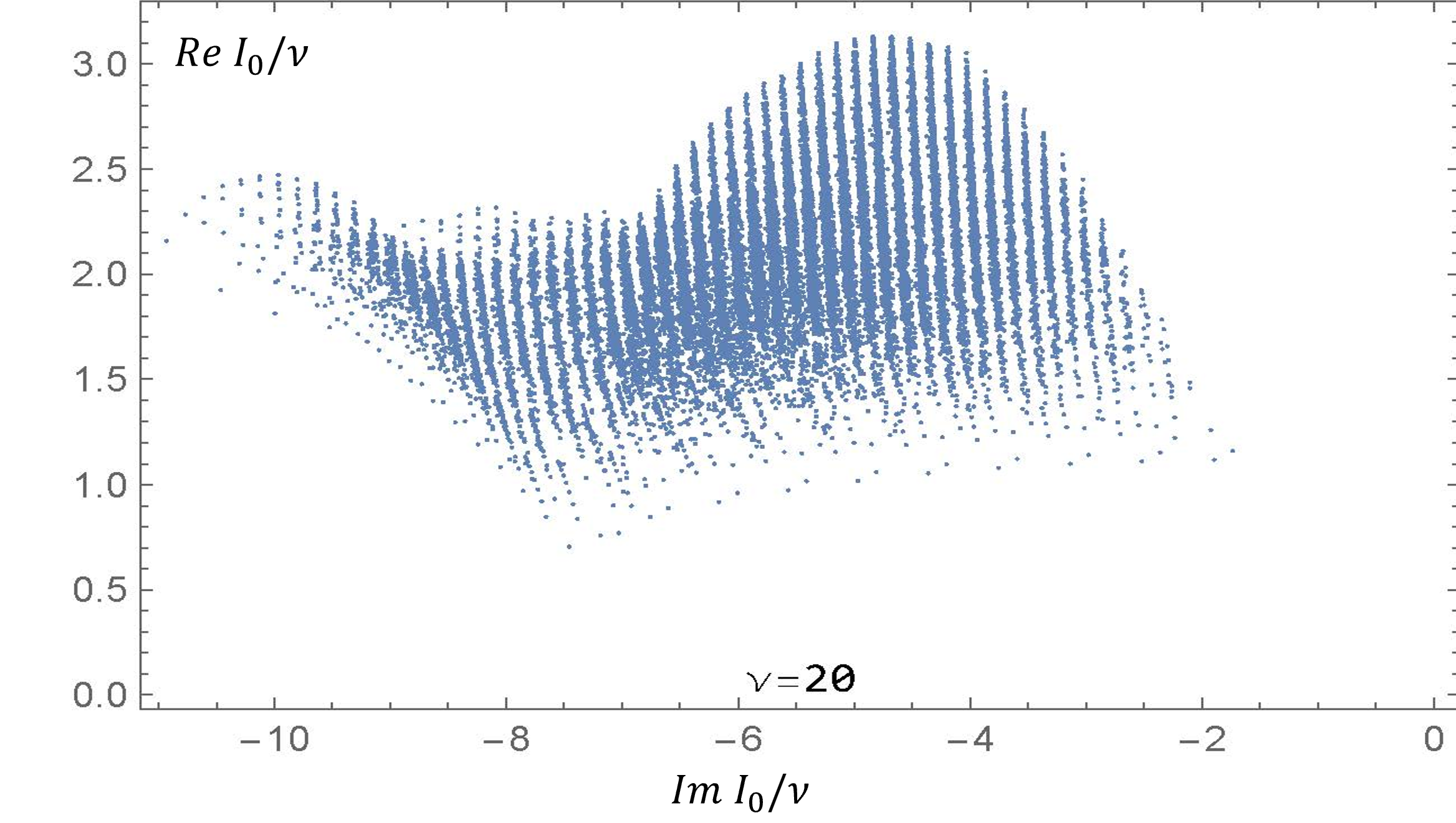}
    \includegraphics[scale=0.21]{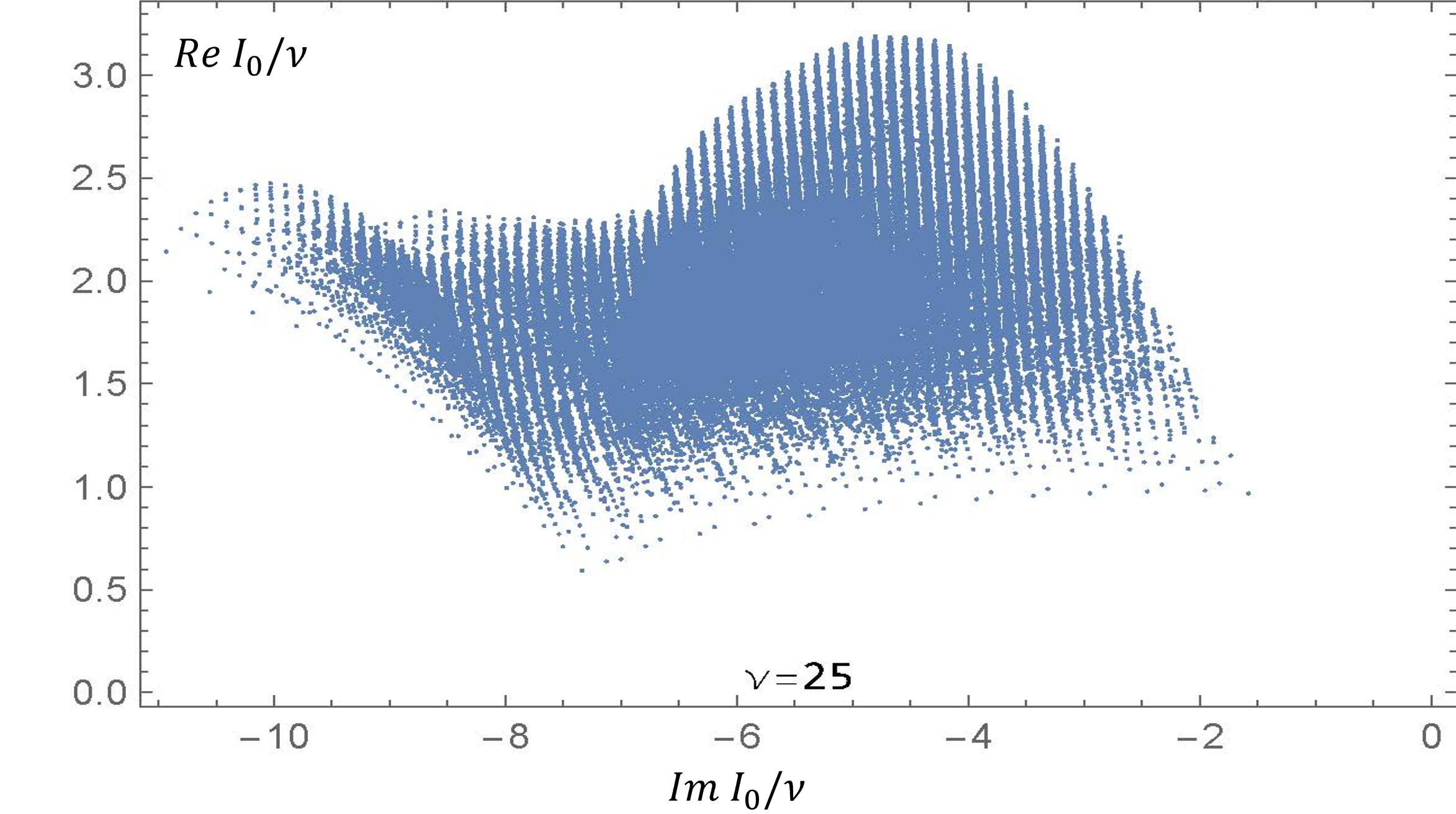}
    \includegraphics[scale=0.4]{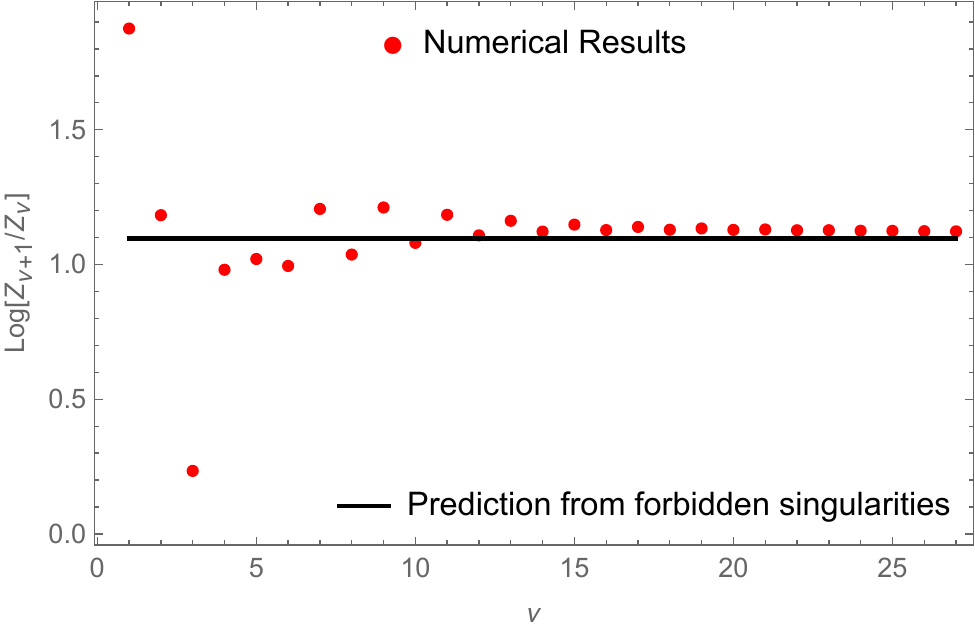}
    \caption{Numerical results for the distribution of the re-scaled complex values for the effective actions $(-\mathcal{I}_0/\nu)$ for all $|Y^1|+|Y^2|=\nu$. The plots converge together with that of $\ln{\left(\bar{\mathcal{Z}}_{\nu+1}/\bar{\mathcal{Z}}_\nu\right)}$ for sufficiently large values of $\nu$. The plots are generated for $\epsilon_H=60,h_L =1$.}
    \label{fig:YD_numerics}
\end{figure}

This can be facilitated by having a glimpse of the global distribution of $\mathcal{I}_0$ over the configuration space. To this end, we plot some numerical results for moderately large $\nu$, so that while universal patterns in the distribution has already formed, it is still possible to enumerate all Young tableaux configurations satisfying $|Y^1|+|Y^2|=\nu$. In particular, we numerically compute the (complex) values of $\mathcal{I}_0(Y^1,Y^2)$ for all $Y^{1,2}$ satisfying $|Y^1|+|Y^2|=\nu$. For each fixed $\nu$, they are plotted as an ensemble of points on the complex plane of $\mathcal{I}_0$. Such ensembles are plotted and compared for a series of $\nu$. We see that the ensembles exhibit a linear-in-$\nu$ scaling in its global distribution. This can be shown by plotting the rescaled ensembles defined by $\mathcal{I}_0(Y^1,Y^2)/\nu$ and observing that, apart from some fine-grained details, the shapes and ranges of the distributions of the rescaled ensembles converge to well-defined limits as $\nu$ increases, see Figure (\ref{fig:YD_numerics}). This happens together with the convergence of $\bar{\mathcal{Z}}_{\nu}/\bar{\mathcal{Z}}_{\nu+1}$ to a limiting value that numerically matches the forbidden singularity $z^*$, indicating that the observed universality is indeed related to the phenomenon of forbidden singularity.    

These evidences suggest that there exists a dominant class of configurations, whose effective actions grow linearly in $\propto \nu$. We need to pin-point this class of configurations. This can be done, for example, by identifying a particular scaling ansatz in large $\nu$: 
\be \label{eq:scaling_def}
Y^\alpha = Y(\nu,y^\alpha),\;\;\alpha=1,2
\ee
such that: 
\be\label{eq:scaling_def_2}
\lim_{\nu \to \infty}\mathcal{I}_0(Y^1,Y^2) = \nu \cdot f(y^1,y^2)
\ee
Here $y^{1,2}$ are the ``atomic" data that define for each $\nu$ a corresponding $Y^{1,2}$ from the dominant class -- the ``DNA" for the growth of $Y^{1,2}$ as $\nu$ increases. At large and fixed $\nu$, the sum over the dominant class of $Y^{1,2}$ can be reduced to that over the atomic structures $y^{1,2}$. More over, having scaled out the large $\nu$ factor, the atomic data are naturally continuous variables, thereby translating the sum into integrals:
\be\label{eq:YD_conjecture}
\sum_{|Y^1|+|Y^2|=\nu} e^{-\mathcal{I}_0(Y^1,Y^2)} \sim \int \mathcal{D}y^{1,2}\; e^{-\nu f(y^1,y^2)}
\ee
where we have denoted the sum (integral) over the atomic data abstractly as $\mathcal{D}y^{1,2}$. 

Effectively, the right-hand side of (\ref{eq:YD_conjecture}) gives a statistical theory for treating the sum over instanton configurations $Y^{1,2}$ at large $\nu$. To make it concrete, the key is to find the scaling ansatz (\ref{eq:scaling_def}) explicitly. To this end, we make the following proposal for (\ref{eq:scaling_def}):
\be\label{eq:YD_scaling}
Y^1= \nu \times \left(y^1_1,y^1_2,....\right),\;\;\;Y^2=\nu \times \left(y^2_1,y^2_2,....\right)
\ee
where $\lbrace y^1_1\geq y^1_2 \geq y^1_3... \rbrace$ and $\lbrace y^2_1\geq y^2_2 \geq y^2_3...\rbrace
$ are sequences of non-increasing non-negative real numbers that satisfy the normalization constraint: 
\be\label{eq:constraint_norm}
\sum^\infty_{X=1} y^1_X + \sum^{\infty}_{X=1} y^2_X = 1
\ee
A finite Young diagram needs to terminate beyond a maximal row index, i.e. $Y_X =0$ for $X\geq X_{\text{max}}$; at $\nu\to \infty$, we have relaxed this condition and only imposed asymptotic termination: $\lim_{X\to \infty}y_{X}=0$. Pictorially, in this ansatz both Young tableaux grow along a particular dimension, i.e. horizontally, as their total size $\nu$ increases. They are string-like objects, but carry non-trivial structures along the transverse direction. A similar scaling of Young tableaux appears in the usual NS limit of Nekrasov partition functions \cite{Bourgine_2013}, in which the ``stretching" is controlled by large $c$ instead of large $\nu$. These are to be contrasted with the typical large $\nu$ behavior of Young diagrams in more conventional statistics, e.g. equal weight, which grows like $\sqrt{\nu}$ along both directions. Such two-dimensional Young tableaux will be discussed in the section (\ref{sec:2d_YD}). 

The preference for the dominant class of Young tableaux to grow horizontally can be traced roughly to the fact that in (\ref{eq:Nekrasov}), the vertical positions ($i,\ell_Y$) of boxes are multiplied by $\epsilon_1 \propto c$ while the horizontal positions ($j,a_Y$) are multiplied by $\epsilon_2\propto \mathcal{O}(1)$. Intuitively this makes the vertical dimension ``heavy" and the horizontal dimension ``light", leading to the highly anisotropic growth in (\ref{eq:YD_scaling}). Similar reasoning was used in the NS limit \cite{Bourgine_2013}. We make a more quantitative argument in the section (\ref{sec:2d_YD}) based on explicit properties of the effective action, by showing that the horizontally growing Young tableaux maximizes the scaling property of the effective action in large $\nu$.     

We now proceed by plugging the scaling ansatz (\ref{eq:YD_scaling}) into the effective action (\ref{eq:effective_action}), and take the leading order terms in large $\nu$. We will refer to this as the large $\nu$ action. At first glance, the expressions in (\ref{eq:effective_action}) may appear to produce many leading-order terms that scale as $ \sim \nu \ln{\nu}$. It is easy to verify that such terms in fact cancel out, and giving rise to a large $\nu$ action that scales linearly with $\nu$, as expected from (\ref{eq:scaling_def_2}):
\bea\label{eq:linear_scaling}
\lim_{\nu \to \infty}\mathcal{I}_0\left(\nu \cdot  y^1, \nu \cdot y^2\right) = \nu \cdot f(y^1,y^2) 
\eea
where $f$ depends only on the atomic Young tableaux $y^{1,2}$. Its explicit form can be easily derived from (\ref{eq:effective_action}) by taking large $Y^{1,2}_X$, and is given by: 
\bea\label{eq:f}
f(y_1,y_2) &=&\sum^2_{\alpha,\beta=1}\sum^\infty_{X=-1} \Delta^{\alpha,\beta}_X \ln \Delta^{\alpha,\beta}_X + \sum^2_{\alpha=1}\sum^\infty_{X=0} \Delta^{\alpha,\alpha}_X G^\alpha_X\nonumber\\
&+&i\pi \tilde{\Theta}
\eea
where we have made a shift of index for $y^2_X \to y^2_{X-1}$ and defined $\Delta^{\alpha,\beta}_X = y^\alpha_X-y^\beta_{X+1}$. We have also extended the index range down to $X=-1$ by defining $y^1_{-1}=y^1_0= y^2_{-1}=0$. The re-scaled phase factor $\tilde{\Theta}$ is now determined by the cross-pattern: 
\bea\label{eq:Theta_rescaled} 
\tilde{\Theta}&=&2y^1_1+2y^2_0+(y^2_0-\max\{y^1_1,y^2_1\})^+\nonumber\\
&+&\sum_{X=1}^{\infty}(\min\{y^1_X,y^2_X\}-\max\{y^1_{X+1},y^2_{X+1}\})^+\nonumber\\
&-&\sum_{X=0}^{\infty}(\Delta^{2,1}_X)^++\sum_{X=0}^{\infty}(-\Delta^{1,2}_{X+1})^+ - \sum^\infty_{X=0} \Delta^{2,1}_X
\eea
An important property of (\ref{eq:Theta_rescaled}) is that for any cross-pattern, $\tilde{\Theta}$ can be reduced to the following form: 
\be\label{eq:phase_property}
\tilde{\Theta} = \sum^{\infty}_{X=1} c_X \cdot y^1_X+\sum^\infty_{X=0} d_X \cdot y^{2}_X
\ee
where $c_X$ and $d_X$ are odd integers. We provide a proof of this in the Appendix (\ref{app:phase_property}). 

The explicit form of the large $\nu$ action (\ref{eq:f}) makes the underlying dynamical structure transparent: it describes two chains represented by $\Delta^{\alpha,\alpha}_X$ interacting through terms involving $\Delta^{\alpha\neq\beta}_X$, and coupled to two external potentials that differ only by a constant term.: 
\be 
G^\alpha_X = \ln{\left[\frac{\Gamma(X+2)\Gamma(2-\alpha-i\alpha_H)}{\Gamma(X+1-i\alpha_H)}\right]}
\ee
The sum over Young tableaux can now be written more explicitly as an infinite dimensional integral: \footnote{We ignore here and later non-trivial integration measures potentially coming from changes of variables. They do not affect the result at the leading order in $\nu$. } 
\bea\label{eq:path-integral}
\bar{\mathcal{Z}}_\nu & \sim & \int \left(\prod^{\infty}_{i=1} dy^1_i\prod^\infty_{i=0} dy^2_i\right)\; \exp{\left(-\nu \cdot f(y^1,y^2)\right)}\nonumber\\
&\times &\delta\left(\nu\sum^2_{\alpha=1}\sum^{\infty}_{X=0} y^\alpha_X -\nu\right)
\eea
In what follows we shall write the infinite dimensional integral over $y^1_X$  and $y^2_X$ simply as $\int\mathcal{D} y^1 \mathcal{D}y^2$. 

\subsection{Saddle-point analysis}
We now evaluate the large $\nu$ integral (\ref{eq:path-integral}) at the leading order in large $\nu$. This can be done by employing the saddle-point approximation. To this end, we add a Lagrange multiplier $\lambda$ to implement the $\delta$-function constraint on the normalization, and write (\ref{eq:path-integral}) as: 
\bea\label{eq:constraint_path_integral}
\bar{\mathcal{Z}}_\nu &\sim & \int \mathcal{D}y^1 \mathcal{D} y^2 d\lambda\; \exp\Bigg(-\nu \Big[\lambda \sum_{\alpha=1}^{2}\sum^\infty_{X=0} y^\alpha_X\nonumber\\
&+& f(y^1,y^2)-\lambda\Big]\Bigg)
\eea
The dynamical variables are $y^{1,2}_X$ and $\lambda$, and the saddle-point equations of motion consist of: 
\be\label{eq:EOM_1}
\frac{\partial f(y^1,y^2)}{\partial y^1_X} = \frac{\partial f(y^1,y^2)}{\partial y^2_X}=-\lambda,\;\;\;\sum_{\alpha=1}^{2}\sum^\infty_{X=0} y^\alpha_X=1 
\ee
The result for the path-integral of $\bar{\mathcal{Z}}_\nu$ at the leading order in large $\nu$ can then be obtained from the on-shell action of (\ref{eq:constraint_path_integral}) by plugging in the solution of (\ref{eq:EOM_1}). A simplifying property of (\ref{eq:constraint_path_integral}) is that $f(y^1,y^2)$ is in fact homogeneous in $y^1$ and $y^2$, as can be easily checked: 
\be 
f(k \cdot y^1, k \cdot y^2) = k \cdot f(y^1,y^2),\;\;\;\forall k\in \mathds{R}
\ee
As a result, the equation of motion with respect to the re-scaling mode $k$ dictates the following identity when evaluated on-shell:
\be
\lambda\sum_{\alpha=1}^{2}\sum^\infty_{X=0} y^\alpha_X + f(y^1,y^2)=0
\ee
The saddle-point approximation for (\ref{eq:constraint_path_integral}) is thus simply given by the on-shell value $\lambda$ of (\ref{eq:EOM_1}) as: 
\be\label{eq:on_shell_1}
\bar{\mathcal{Z}}_\nu \sim e^{\nu \lambda}
\ee

Looking back at the numerical plots in Figure (\ref{fig:YD_numerics}), we see that the ensemble of points does not feature a particular configuration that dominates and directly contributes $\bar{\mathcal{Z}}_\nu$ at large $\nu$ -- in fact, the ensemble contains a large number of points whose contributions are much larger in magnitude than that of $\bar{\mathcal{Z}}_\nu$. Very delicate cancellations occur among these contributions. We take this as evidence that the saddle-point of (\ref{eq:YD_conjecture}), if present, does not occur as realistic Young tableaux, but rather as their analytic continuations. We can probe this possibility now because the degrees of freedom are positive real numbers $y^{1,2}_X \in \mathds{R}^+$ instead of the original positive integers $Y^{1,2}_i\in \mathds{N}$, allowing us to deform the integration contour and look for complex saddle-points $y^{1,2}_X \in \mathds{C}$. 

In doing this, in principle we need to keep track of how the integration contour is deformed. A rigorous prescription for defining the integration contour of $\int\mathcal{D}y^1 \mathcal{D} y^2$ on $\mathds{R}^+$ involves additional topological constraints. There are two classes of constraints: the sequences $y^{1,2}_X$ are ordered $y^{1,2}_X\geq y^{1,2}_{X+1}$; when projected onto a sector with fixed expression for the phase index $\tilde{\Theta}$, $y^1$ and $y^2$ also need to satisfy a particular cross-pattern. It is a subtle to keep track of these constraints once $y^{1,2}_X$ move away from $\mathds{R}^+$. In principle we can proceed by fixing an ordering of the dynamic variables consistent with the constraint, and consider the path-integral as a sequence of real-valued single-variable definite integrals, e.g:  
\be\label{eq:YD_cone}
\int_{\lbrace y^1_X \geq y^1_{X+1}\rbrace} \mathcal{D}y^1 =  \int^\infty_0 dy^1_1 \int^{y^1_1}_0 dy^1_2 \int^{y^1_2}_0 dy^1_3 ...
\ee
This defines the interior of a cone in real space. To construct an analytically continuation of (\ref{eq:YD_cone}), we embed this cone in the complex space $\prod_X \lbrace z^1_X \in \mathds{C}\rbrace$ as a real sub-manifold with boundaries. One can then deform this real sub-manifold, e.g, by deforming the integration contour, as well as the movable end-points, away from the real-axis: 
\be
\int^\infty_0 dy^1_1 \int^{y^1_1}_0 dy^1_2 ... \to \int_{\mathcal{C}^{\infty}_0} dz^1_1 \int_{\mathcal{C}^{z^1_1}_0} dz^1_2 ...
\ee
where $\mathcal{C}^z_0$ denotes a deformed contour in the complex plane that connects $0$ and $z$. Notice that the action (\ref{eq:f}) features only mild (integrable) singularities of the form $\lim_{x\to 0}x\ln x$, which occur only at the boundary of the cone. Intuitively this makes the path-integral well-behaved under the
analytic continuation just described. Motivated by these, in this work we assume that there is no subtle obstruction for performing the required analytic continuations to reach the complex saddle-points we consider.  

We now take into account the explicit form of $f(y^1,y^2)$ and write down the equations of motion:
\bea\label{eq:EOM_2}
\frac{\Delta^{\alpha,\alpha}_X \Delta^{\alpha,\beta}_X}{\Delta^{\alpha,\alpha}_{X-1}\Delta^{\beta,\alpha}_{X-1}}=e^{-\lambda} \left(\frac{i\alpha_H-X}{X+1}\right),\;\;\alpha \neq \beta
\eea
Due to the asymmetric initial conditions imposed on $y^{1,2}_X$, the row-indices of Eq (\ref{eq:EOM_2}) begin at $X=1$ for $(\alpha=1,\beta=2)$ and at $X=0$ for $(\alpha=2,\beta=1)$. We also mention that (\ref{eq:EOM_2}) is valid for any cross-pattern between $y^1$ and $y^2$. This is due to the general form (\ref{eq:phase_property}) of $\tilde{\Theta}$, which invariably contribute a factor of $(-1)^{c_X, d_X}=(-1)$ when taken variation w.r.t. any $y^{1,2}_X$.  

Eq (\ref{eq:EOM_2}) is a system of coupled non-linear difference equations for $y^{1,2}_{X}$. A closer examination reveals that Eq (\ref{eq:EOM_2}) by itself is under-determined, a general solution depends on three free parameters, for which we can choose as $(y^1_1, y^2_0, \lambda)$.  
This is the right number of free parameters to satisfy the constraints imposed on the solutions:
\be\label{eq:bc}
\sum^2_{\alpha=1}\sum^\infty_{X=0} y^\alpha_X =1,\;\;
\lim_{X\to \infty} y^1_X =0,\;\;\lim_{X\to \infty} y^2_X =0
\ee
We remark that latter two constraints come from the geometric characters of real Young tableaux $y^{1,2}_X\in \mathds{R}^+$. Upon analytic continuation into the complex plane, they survive as well-defined analytic statements even though the original geometrical meanings are lost, so keeping them is a reasonable prescription for saddle-point analysis on the complex-plane. The system of Eq (\ref{eq:EOM_2}) and Eq (\ref{eq:bc}) combined is therefore well-defined with no free parameters.  

Solving the saddle-point equations in full generality remains a difficult task. For example, the algebraic complexity of recursively solving for $y^{1,2}_X$ beginning from small $X$ grows quickly with $X$. To make progress, we simply propose a special property for the saddle-point solution:
\be\label{eq:guess_1}
y^2_{X}=0,\;\;\;X\geq 0
\ee
This proposal trivially satisfies the asymptotic boundary condition $\lim_{X\to\infty}y^2_X =0$. It also solves automatically the subset of Eq (\ref{eq:EOM_2}) with $(\alpha=2,\beta=1)$: 
\be \label{eq:EOM_3}
\Delta^{2,2}_{X}\Delta^{2,1}_X = e^{-\lambda}\left(\frac{i\alpha_H-X}{X+1}\right)\Delta^{2,2}_{X-1}\Delta^{1,2}_{X-1}
\ee
To see this, we begin with the special case of $X=0$, expanded in the form: 
\be 
(y^2_0-y^2_1)(y^2_0-y^1_1) = ie^{-\lambda}\alpha_H \left(y^2_0 \right)^2
\ee
This is satisfied by setting $y^2_0 = y^2_1=0$. The vanishing of $y^2_X$ can then be propagated to large index $X$ by requiring $\Delta^{2,2}_{X}=y^2_X-y^2_{X+1}=0$, which satisfies (\ref{eq:EOM_3}) for finite $\lambda$. 

The remaining equations in (\ref{eq:EOM_2}) now simplifies to:
\bea\label{eq:EOM_5}
\frac{\Delta^{1,1}_X}{\Delta^{1,1}_{X-1}}=e^{-\lambda} \left(\frac{X-i\alpha_H}{X+1}\right),\;X\geq 1
\eea
with the constraints: 
\be
\sum^\infty_{X=1} y^1_X = 1,\;\;\;\; \lim_{X\to \infty} y^1_X = 0
\ee
From (\ref{eq:EOM_5}) it is easy to first solve:
\be
\Delta^{1,1}_X=\frac{k\cdot\Gamma(i\alpha_H)(-1)^X e^{-\lambda X}}{\Gamma(X+2)\Gamma(i\alpha_H-X)}
\ee
where we have set the initial value $y^1_1 = k$. We can then organize the $\Gamma$-function factors in terms of the binomial coefficients $\binom{i\alpha_H-1}{j}$, and compute $y^1_X = y^1_1+ \sum^{X-1}_{j=1}\Delta^{1,1}_j$: 
\bea\label{eq:saddle}
y^1_X &=& k \sum_{j=0}^{X-1}\binom{i\alpha_H-1}{j}(j+1)^{-1}(-1)^je^{-j\lambda}\nonumber\\
&=& k\; e^{\lambda} \sum_{j=0}^{X-1}\binom{i\alpha_H-1}{j} \int_0^{e^{-\lambda}} dy\; (-y)^j 
\eea
We see that the saddle-point solution (\ref{eq:saddle}) for $y^1_x$ does takes values in the complex plane, as was expected.  From (\ref{eq:saddle}) it is easy to compute the asymptotic value $y^1_{\infty}$ by re-summing the infinite series, for which we can switch the order of summation and integration and obtain:
\bea
    y^1_{\infty}&=& k e^{\lambda} \int^{e^{-\lambda}}_0 dy\; \sum^{\infty}_{j=0} \binom{i\alpha_H-1}{j} (-y)^{j} \nonumber\\
    &=& k e^{\lambda} \int^{e^{-\lambda}}_0 dy (1-y)^{i\alpha_H-1}\nonumber\\
    &=& \frac{ik e^\lambda}{\alpha_H}\left[\left(1-e^{-\lambda}\right)^{i\alpha_H}-1\right]
\eea
By imposing the boundary condition $y^1_{\infty}\to 0$, we can then solve the on-shell values of $\lambda$:
\be\label{eq:lambda_onshell}
e^{-\lambda_n} = 1-e^{-\frac{2\pi n}{\alpha_H}},\;\;\;n \in \mathds{Z}
\ee
There exists infinitely many on-shell values for $\lambda_n$, each gives a potential contribution $e^{\lambda_n \nu}$ to $\bar{\mathcal{Z}}_\nu$. One can easily verify that the most dominant contribution comes from that of $n=1$: 
\be
\lim_{\nu\to \infty}\bar{\mathcal{Z}}_\nu \sim \left(1-e^{-\frac{2\pi}{\alpha_H}}\right)^{-\nu}
\ee
We therefore extracts from (\ref{eq:f_scaling}) a radius of convergence that matches the forbidden singularity:
\be 
r=1-e^{-\frac{2\pi}{\alpha_H}} = z^*
\ee
We notice that the saddle-point analysis captures not only $z^*=z_1$ that shows up in the vacuum block, but also the remaining forbidden singularities of $z_{n}$ that are relevant to the un-physical blocks.  

We emphasize that since the equations of motion (\ref{eq:EOM_2}) take the same form for any possible expression (\ref{eq:phase_property}) of $\tilde{\Theta}$, the validity of our saddle-point (\ref{eq:guess_1},\ref{eq:saddle}) as a solution to the equations of motion is independent of the crossing-pattern between $y^1$ and $y^2$. 

For completeness, we also work out the normalization factor $k$ by summing over $y^1_x$  using the result of (\ref{eq:saddle}) and plugging the on-shell value (\ref{eq:lambda_onshell}): 
\bea
&& \sum^{\infty}_{X=1} y^1_X =  k e^{\lambda} \sum^\infty_{X=1} \sum_{j=0}^{X-1}\binom{i\alpha_H-1}{j} \int_0^{e^{-\lambda}} dy\; (-y)^j \nonumber\\
&=& ke^{\lambda} \int^{e^{-\lambda}}_{0} dy\sum^\infty_{j=0} \binom{i\alpha_H-1}{j} \left(\sum^\infty_{i=1}  1- \sum^j_{i=1} 1\right) (-y)^j\nonumber\\
&=& k e^{\lambda} \int^{e^{-\lambda}}_0 dy \sum^{\infty}_{j=0} \binom{i\alpha_H-1}{j} (-j) (-y)^j \nonumber\\
&=& k (e^{-\lambda}-1)^{-1} 
\eea
where going from the second to the third line we have used that the factor $\left(\sum^\infty_{i=1} 1\right)$ is proportional to $(1-e^{-\lambda})^{i\alpha_H}-1=0$. Imposing the normalization condition then gives that:
\be
k=e^{-\lambda} -1= -e^{-\frac{2\pi}{\alpha_H}}
\ee

In summary, we found the dominant (complex) saddle-point (\ref{eq:guess_1}, \ref{eq:saddle}) that controls the leading-order behavior of the instanton gas described by the effective action (\ref{eq:effective_action}) at large $\nu$. Based on this, we identified a critical fugacity $r$ for the effective fugacity expansion in terms of its radius of convergence, and found that it matches exactly the forbidden singularity $z^*$ that arise in the heavy-light virasoro vacuum block.  

From these results, we have recast the phenomenon of forbidden singularity explicitly as a phase transition in eigenstates, via the AGT correspondence. In the context of the equivalent instanton gas description, this phase transition is an instability of instanton proliferation $\nu\to \infty$ at the critical fugacity. The proliferation occurs in a particular way that is captured by $Y^{1,2}_X =\nu \cdot y^{1,2}_X$ through the saddle-points for $y^{1,2}_X$ in (\ref{eq:guess_1}, \ref{eq:saddle}). The saddle-point is characterized by  $y^2_X=0$ -- or more precisely by microscopic $Y^2$ at large $\nu$, and by complex $y^1_X$. We emphasize that the complex nature means that the saddle-point describes a collective behavior, rather than a realistic configuration that dominates over others, of the instanton gas ensemble. This is consistent with what we observed in the numerical results.        
 
\subsection{Additional solutions}
The saddle-point solution we found at large $\nu$ correctly gives the radius of convergence dictated by the forbidden singularity. However, we remark that the way to solve the equations of motion (\ref{eq:EOM_2}) is by no means unique, there appears to be a very rich landscape of additional solutions. As was argued, the number of free parameters (three) for solving (\ref{eq:EOM_2}) matches the number of conditions imposed, additional saddle-point solutions exist as isolated points. While it is difficult to exhaust all possible solutions, we mention a few more examples to get a glimpse of what lies beyond the one we found. 

To this end, we remind that the equations of motion (\ref{eq:EOM_2}) is invariant under the symmetry $y^1_X \leftrightarrow y^2_X$ that can be performed independently for any $X\geq 1$. It is only broken at $X=0$ by the distinct initial conditions imposed on $y^1_0$ and $y^2_0$. This is a vast number of discrete symmetries. Such systems often dynamically prefer saddle-points that exhibit sharp symmetric properties -- either symmetric or maximally symmetry-breaking. We focus on these solutions. 

In fact, the saddle-point (\ref{eq:guess_1},\ref{eq:saddle}) we have found is an example of a maximally symmetry-breaking solution. Formally by performing the switching $y^1_X \leftrightarrow y^2_X$ for any subset of $X\geq 1$, we obtain additional solutions with the same on-shell action. The is analogous to the scenario of the spontaneous breaking of the $y^1 \leftrightarrow y^2$ symmetry, in which the degenerate ground-states appear that transform into one another under the symmetry. On the other hand, from the definition of the original integration contour (\ref{eq:YD_cone}) we see that setting any $y^2_X=0$ makes the volume of the sub-cone vanish: 
\be
\lim_{y^2_X\to 0} \text{Vol} \left(\int^{y^2_X}_0 dy^1_{X+1} \int^{dy^1_{X+1}}_0 dy^2_{X+1} ...\right) \to 0
\ee
So for any analytic continuation, this localizes $y^2_{X'}=0$ for $X'\geq X$. Based on this, we discard the additional solutions obtained by performing the switching, because the initial condition $y^2_0=0$ remains fixed for all these solutions.

There exists another class of maximally symmetry-breaking solutions. They can be obtained by assuming $y^1_X=0$ but now starting with $y^2_0\neq 0$. We find a solution of similar form to (\ref{eq:guess_1}, \ref{eq:saddle}): 
\be\label{eq:saddle_3}
y^1_X = 0,\;\;\;y^2_X \propto  \sum^X_{j=0} (-1)^j e^{-\lambda j} \binom{i\alpha_H}{j}
\ee
Additional solutions can be obtained from (\ref{eq:saddle_3}) via $y^1_X \leftrightarrow y^2_X$ for any subset of $X\geq 1$. However, imposing the asymptotic boundary condition $\lim_{X\to \infty} y^2_X=0$ on this solution gives the on-shell value:
\be 
\lim_{X\to \infty}y^2_X \propto (1-e^{-\lambda})^{i\alpha_H} =0 \to \lambda =0
\ee
It gives a contribution to $\bar{\mathcal{Z}}_\nu$ that is sub-dominant at large $\nu$ compared to the saddle-point (\ref{eq:guess_1},\ref{eq:saddle}). 

\begin{figure}[!htp]
\centering
\includegraphics[scale=0.65]{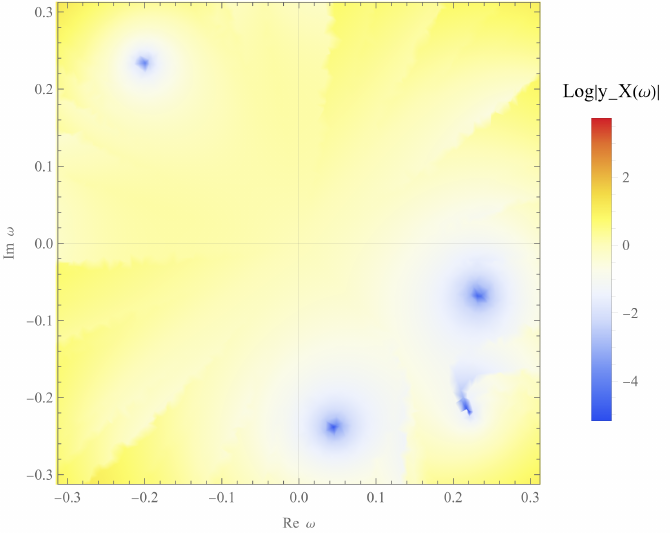}
\caption{The roots $\omega_{sym}$ (blue dots) of the symmetric solution \(\lim_{X\to \infty}y_X(\omega)\) in the complex \(\omega\)-plane for \(\alpha_H=10\), truncated at the order \(X_{\mathrm{max}}=120\). In this case, several complex $\omega_{sym}$ are found with smaller modulus than the corresponding $\sqrt{z^*}\sim 0.683$. They all give on-shell values of $\lambda_{sym}$ satisfying $\text{Re}\; \lambda_{sym}> \ln{(1/z^*)}$. }\label{fig:symmetric solution}
\end{figure}

It is also possible to solve the equations of motion (\ref{eq:EOM_2}) by assuming the solution to be symmetric: $y^1_X = y^2_X=y_X$ for $X \geq 1$.  In this case we can obtain: 
\bea\label{eq:saddle_4}
y_X(\omega) \propto &&\sum_{j=2}^{X}\left[\left(1-\frac{1}{\omega\sqrt{i\alpha_H}}\right)\omega^{2j}\binom{i\alpha_H}{j}\right]^{\frac{1}{2}}\nonumber \\
&+&\left(\omega\sqrt{i\alpha_H}-1\right)
\eea
Unfortunately, the asymptotic value $\lim_{\infty} y_X(\omega)$ now involves an infinite power series in $\omega =e^{-\lambda/2}$ that cannot be re-summed into closed-form. Consequently we cannot find a closed-form expression for the on-shell value of $\lambda$ that solves $\lim_{X\to \infty} y_X(\omega)=0$. Instead we numerically solve $\lambda$ for specified values of $\alpha_H$, by truncating the series too sufficiently high order that converges well near its roots. We find that the symmetric solutions can yield roots $\omega_{sym}=e^{-\lambda_{sym}/2}$ whose on-shell values of $\text{Re}\;\lambda_{sym}$ are larger than $\lambda = \ln{(1/z^*)}$ of the saddle-point (\ref{eq:guess_1}, \ref{eq:saddle}), thus corresponding to a more dominant contribution $e^{\nu\cdot\lambda_{sym}}$ to $\bar{\mathcal{Z}}_\nu$, see Figure (\ref{fig:symmetric solution}) for an illustration of the roots for $\alpha_H=10$.  

The fact that the symmetric solution can give a more dominant contribution than (\ref{eq:guess_1}, \ref{eq:saddle}) appears to suggest a smaller radius of convergence $r<z^*$. This is in conflict with the CFT prediction of $r=z^*$, which is verifiable numerically at large but finite $c$. It therefore must be the case that the symmetric solution as a saddle-point does not contribute to the path-integral (\ref{eq:path-integral}), because its Lefschetz thimble in the configuration space of $y^{1,2}$ do not intersect with the defining integration contour of (\ref{eq:path-integral}). To understand why this can be the case, we remind that a saddle-point contribution to $\bar{\mathcal{Z}}_\nu$ of the form $e^{\nu \lambda}$ is associated with a fugacity singularity of $\mathcal{Z}^{\text{U(2)}}_{\text{inst}}(z)$ at $z=e^{-\lambda}$. Recall further that in the semi-classical limit $c\to \infty$, the free energy $\mathcal{F}(z)$ is a multi-valued function on the complex plane, connecting the vacuum block with additional un-physical blocks. It is possible that the singularity of $z$ associated with the symmetric solution does not lie on the principal branch of $\mathcal{F}(z)$. Instead it is located on other branches that are accessible only by passing branch-cuts on the principal branch, e.g. those emanating from the singularity at $z^*$. It therefore does not control the radius of convergence for the series expansion despite its apparent smaller modulus. The branch-cut thus generates an obstruction mechanism for the Lefschetz thimble and the defining integration contour to touch in the configuration space. It is a fascinating future problem to understand this obstruction in more details, and how the symmetric solution contribution arises in the semi-classical CFT computations, e.g. the monodromy method.  

\subsection{Two-dimensional Young tableaux}\label{sec:2d_YD}
We end this section by discussing the contributions from the other class of large Young tableaux that may appear -- those that grow in both directions, i.e. $Y_i \sim X_j \sim \sqrt{\nu}$. We refer to these as two-dimensional Young tableaux. In the large $\nu$ limit, they can be obtained by performing the following scaling operations on two continuous functions $y^{1,2}(x)$ delineating the so-called limiting shapes: 
\be\label{eq:2d_scaling}
Y^{1,2}_X = \sqrt{\nu} \cdot \tilde{y}^{1,2}(X/\sqrt{\nu})
\ee
Since the row-lengths $Y^{1,2}_X$ are still large in (\ref{eq:2d_scaling}), the large $\nu$ action is equally applicable to (\ref{eq:2d_scaling}), now with a modified pre-factor $\nu \to \sqrt{\nu}$: 
\bea\label{eq:2d_action}
&&\mathcal{I}_0\left(Y^1,Y^2\right) \sim \sqrt{\nu}\cdot f(y^1,y^2)\nonumber\\
&&y^{1,2}_X = \tilde{y}^{1,2}(x),\;\;x=X/\sqrt{\nu}
\eea
For (\ref{eq:2d_scaling}) we can further scale out a $\sqrt{\nu}$ factor from the discrete sum over $X$, and write it as an integral over the continuous variable $x$: $\sum_X \to \sqrt{\nu} \int dx$. This could then combine with the $\sqrt{\nu}$ pre-factor to give $\mathcal{I}_0\propto \nu$. It is in fact not true. In the continuous limit, by re-grouping terms the summand of (\ref{eq:f}) can be written in terms the derivatives $\dot{\tilde{y}}^{\alpha}(x) = d \tilde{y}^{\alpha}(x)/dx $ as: 
\bea\label{eq:action_2d_scaling}
-\nu^{-1/2}\sum^{2}_{\alpha=1}\dot{\tilde{y}}^\alpha(x) \ln{\left(\dot{\tilde{y}}^\alpha(x) \tilde{\Delta}(x)\tilde{G}^\alpha(x)\right)} 
\eea
where we have defined: 
\be
\tilde{\Delta}(x) = \tilde{y}^1(x)-\tilde{y}^2(x),\;\;\tilde{G}^{\alpha}(x)= G^\alpha\left(\left[x \sqrt{\nu}\right]\right)
\ee
The $\nu^{-1/2}$ pre-factor in (\ref{eq:action_2d_scaling}) comes from trading finite difference with derivative: $(...)_{X+1}-(...)_X = \nu^{-1/2} \partial_x (....)$. This cancels with the $\sqrt{\nu}$ factor from the sum, making $f(y^1,y^2)$ independent of $\nu$. We therefore see that for two-dimensional Young tableaux of the form (\ref{eq:2d_scaling}), the effective action is: 
\be\label{eq:2d_action_scaling}
\mathcal{I}_0(Y^1,Y^2) \sim \nu^{1/2}
\ee
The two-dimensional Young tableaux can therefore be consistently neglected at $c=\infty$. In fact, applying the same argument to more general two-dimensional Young tableaux with anisotropic scalings: 
\be
Y^{1,2}_X = \nu^{p}\cdot \tilde{y}^{1,2}(X/\nu^{q}) 
\ee
for $p,q>0$ satisfying $p+q=1$, one can show that in this case the effective action scales as: 
\be
\mathcal{I}_0(Y^1,Y^2) \sim \nu^{p}
\ee
This then proves that the class of string-like Young tableaux (\ref{eq:YD_scaling}), corresponding to $p=1$, maximizes the large $\nu$-scaling. 

\section{High fugacity phase}\label{sec:deconf}
So far we have been focusing on the low fugacity phase where the effective expansion (\ref{eq:eff_fugacity_exp}) at $c=\infty$ is convergent. This phase is independent of $c$ and thus resembles a confined phase. From a finite $c$ point of view, this phase is dominated by instanton gas configurations with $\nu\ll c$, and the proliferation at the critical fugacity $z=z^*$ is a signal to exit this regime and entering that of $\nu\gtrsim c$. We refer to this as the high fugacity phase. It is beyond the scope of this work to perform a complete analysis -- we leave this to the future. Here, we zoom into the special regime of $\nu \gg c$, and perform analysis at the leading order in both large $\nu/c$ and large $c$.  

We first discuss the class of Young tableaux that dominates this regime. The numerical option of exhausting all Young tableaux configurations at a given $\nu$ and observe the distribution of their actions is no longer available for $\nu \gg c$. Instead, we shall restrict to comparing two typical class of large Young tableaux: the string-like type, and the two-dimensional type.

\begin{figure}[h]
\centering
\includegraphics[scale=0.25]{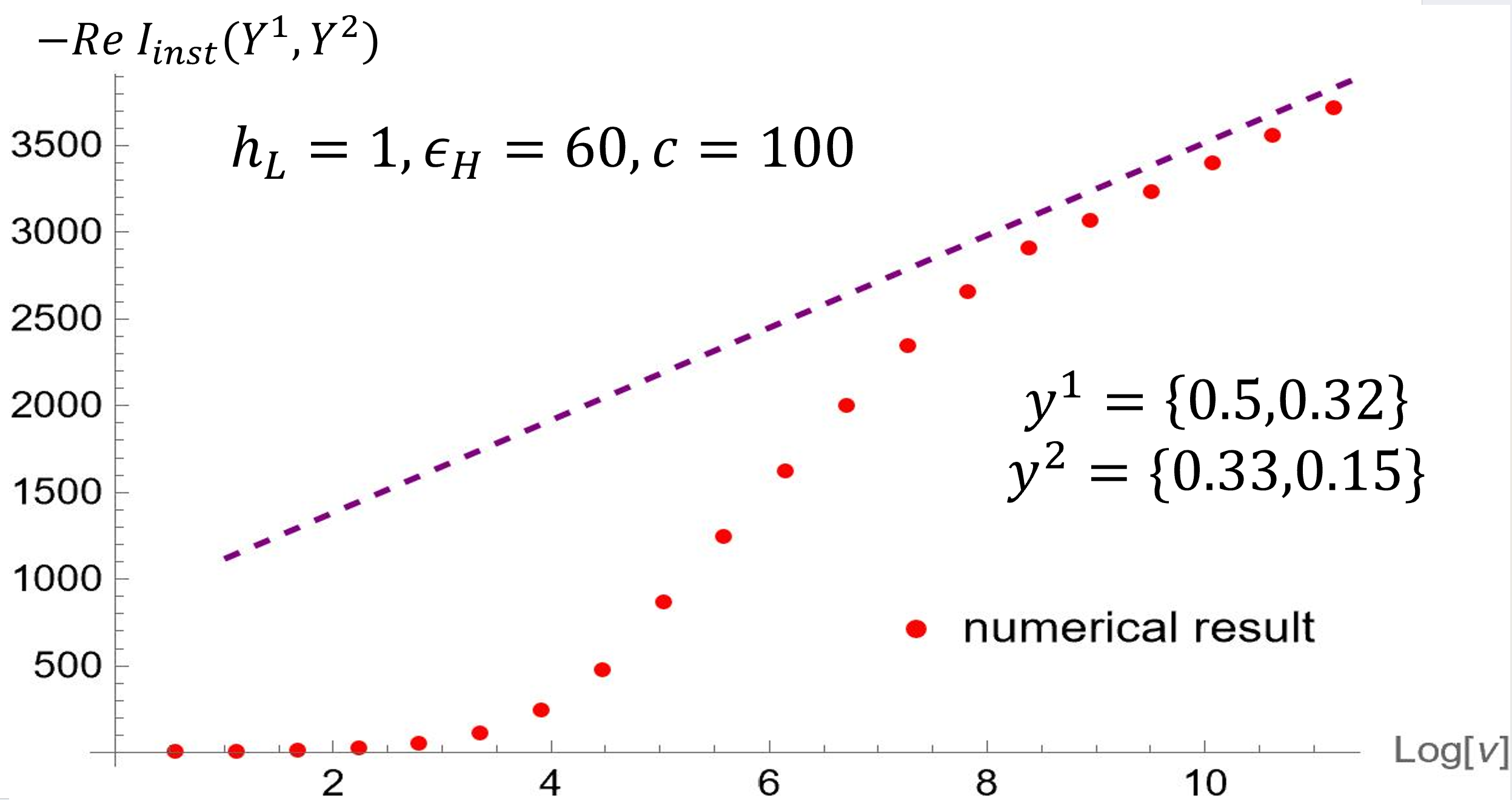}
\caption{Numerical plot (red dots) of the exact action $(- \text{Re}\tilde{\mathcal{I}}_{\text{inst}})$ evaluated at $c=100, \epsilon_H=60, h_L=1$, as a function of $\ln{\nu}$ for stringy Young tableaux $Y^{1,2}=\nu \times y^{1,2}$ with $y^1=\lbrace0.5,0.32\rbrace,\;y^2=\lbrace 0.33,0.15 \rbrace$. Its asymptotic slope in $\ln{\nu}$ matches that of the analytic expression (blue dashed) according to (\ref{eq:eff_deconf}).   }\label{fig:stringYD_large_nu}
\end{figure}

We begin with the stringy type, now with fixed numbers of rows:  $Y^{1,2}_X = \nu \cdot y^{1,2}_X,\;X=1,...,n_{1,2}$. At the leading order in large $\nu/c$ and $c$, we can derive the following form for the effective action of $Y^{1,2}$:
\bea\label{eq:eff_deconf}
\tilde{\mathcal{I}}_{\text{eff}} =  \frac{c}{6}\ln{\left(\frac{\nu}{c}\right)}\bar{n}(i\alpha_H-\bar{n})  + \frac{c}{6} \tilde{f}(y^1,y^2) + ... 
\eea
where $\bar{n}=n_1+n_2$ is the total number of rows of the Young tableaux $Y_{1,2}$, and ... denotes sub-leading terms in large $\nu/c$ and large $c$. See Figure (\ref{fig:stringYD_large_nu}) for a numerical plot of $-\tilde{\mathcal{I}}_{\text{eff}}$ illustrating the transition from the $\propto \nu$ scaling for $1\ll \nu \ll c$ according to (\ref{eq:f_scaling}), to the $\propto \ln{\nu}$ scaling for $\nu \gg c\gg 1$ according to (\ref{eq:eff_deconf}). The explicit form of $\tilde{f}(y^1,y^2)$, as well as the derivation of (\ref{eq:eff_deconf}), can be found in the Appendix (\ref{app:large_nu}). We shall not quote it here. The important part of (\ref{eq:eff_deconf}) is the general scaling  $\bar{\mathcal{I}}_{\text{eff}} \sim c\ln{\nu}$ for the action of the stringy type.

\begin{figure}[h]
\includegraphics[scale=0.25]{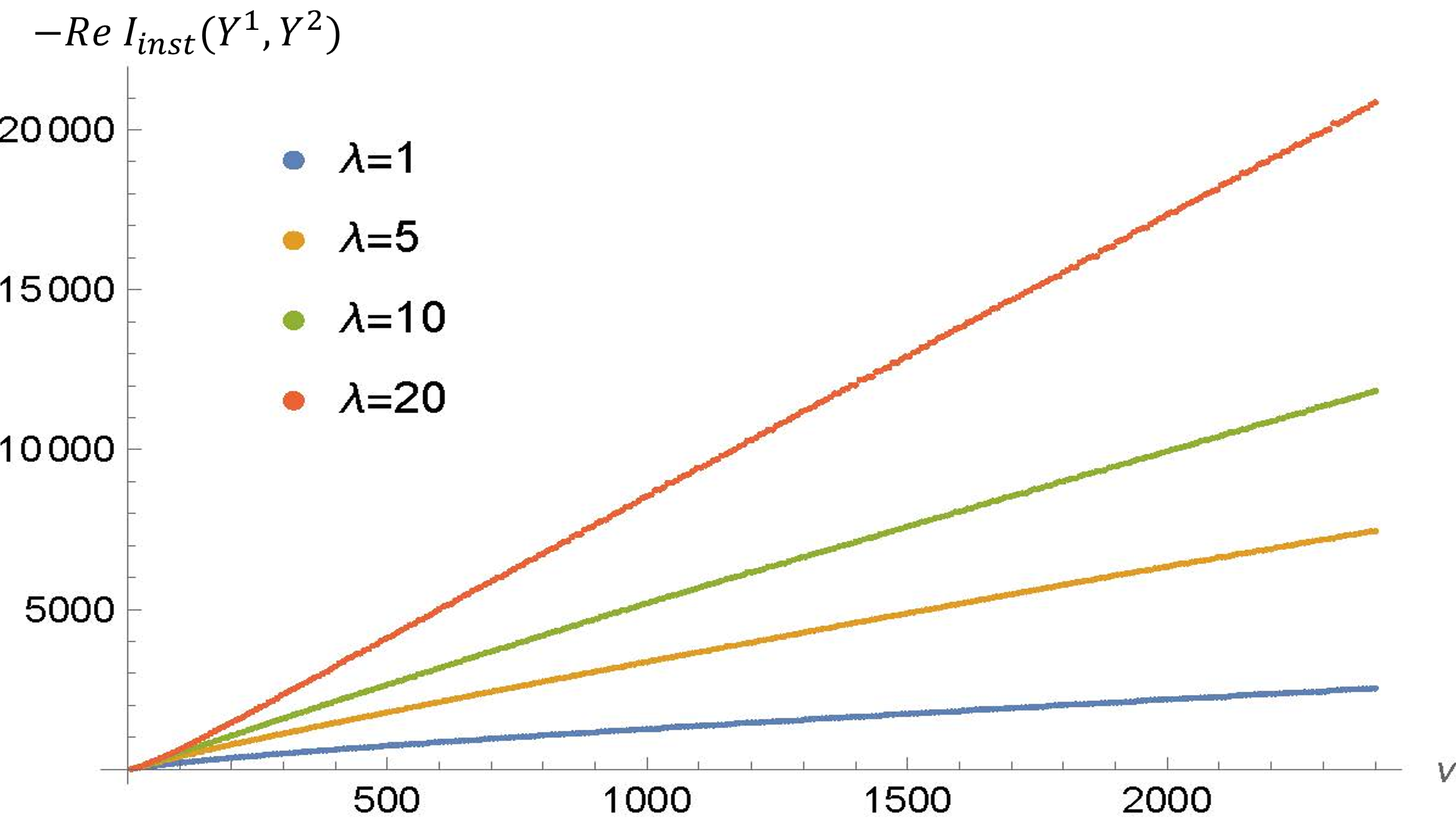}
\includegraphics[scale=0.25]{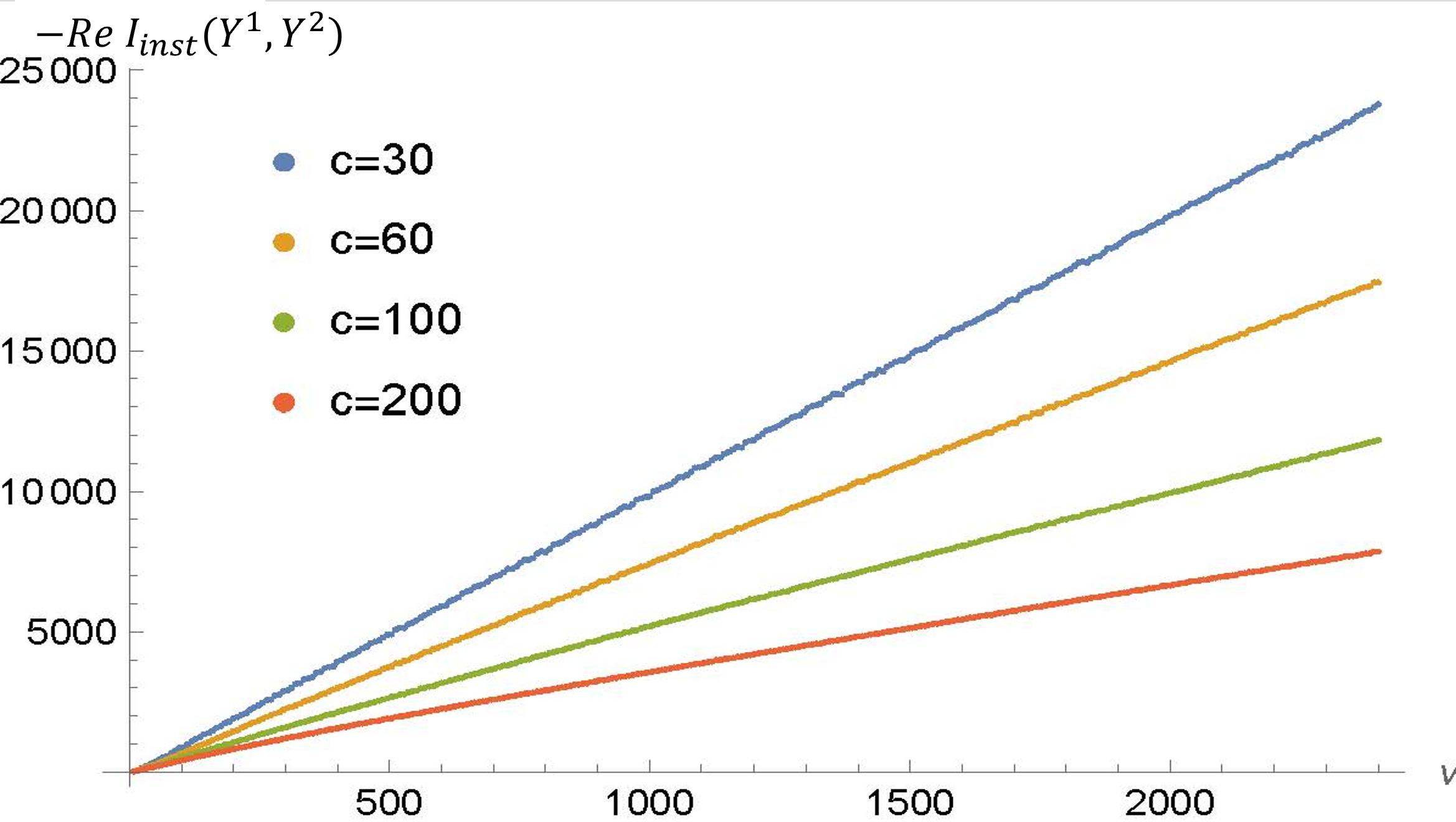}
\caption{Numerical plots for the minus of the real part of the exact actions $(-\text{Re}\mathcal{I}_{\text{inst}})$ as functions $\nu$ for skewed two-dimensional YD (\ref{eq:2dYD_skew}). Left: at fixed $c=100$ for various skew parameter $\lambda$. Right: at fixed skew parameter $\lambda =10$ for various $c$. In all plots we have fixed the limiting curves as $y^{1,2}(x)=\sigma_{1,2}^{-1}\exp{\left(-x^2/\sigma_{1,2}^2\right)},\;\sigma_1=1,\sigma_2=2$, and the remaining parameters as $\epsilon_H=60,h_L=1$.}\label{fig:2d_YD}
\end{figure}

For the two-dimensional type, we extract the scaling behavior of their actions at large $\nu/c$ and $c$ by numerical observations, and leave a more thorough analysis for future investigations. We summarize the results as follows: 
\begin{itemize}
\item For a sequence of $Y^{1,2}_X = \sqrt{\nu} \times y^{1,2}(X/\sqrt{\nu})$ defined by fixed limiting shapes $y^{1,2}(x)$, the scaling behavior of their actions $\mathcal{I}_{\text{inst}}$ transits from $\propto \sqrt{\nu}$ in (\ref{eq:2d_action_scaling}) to $\propto \nu$ with a negative slope for $\nu \gg c$. 
\item There exists a ``run-away" mode that skew the Young tableaux, which we parametrize as $\lambda$: 
\be\label{eq:2dYD_skew}
Y^{1,2}_X(\lambda) = \sqrt{\nu \lambda} \times y^{1,2}\left(X/\sqrt{\nu \lambda^{-1}}\right)
\ee
It is observed that the action scales linearly with $\lambda$: $\mathcal{I}_{\text{inst}} \to \lambda \mathcal{I}_{\text{inst}}$. So the action becomes more negative, thus more dominant, under increasing $\lambda$. Under extrapolation, we may expect that the end point of this run-away direction is indeed the stringy Young tableaux, when $\lambda \sim \nu$. 

\item For a sequence of two-dimensional $Y^{1,2}_X$ with fixed limiting shapes $y^{1,2}(x)$, by varying $c \gg 1$, it is also observed that the slope of the linear regime scales inversely with $c$: $\mathcal{I}_{\text{inst}}\propto c^{-1} \times \nu$.
\end{itemize}
In Figure (\ref{fig:2d_YD}) we illustrate these observations by plotting the action of two-dimensional Young tableaux (\ref{eq:2d_scaling}) for a particular choice of $y^{1,2}(x)$, for a variety of values for $\lambda$ and $c$. Based on these observations, particularly the existence of a run-away channel towards the string-like limit, and the scaling behavior with large $c$, we conclude that the two-dimensional Young tableaux remain irrelevant at large $\nu/c$ and large  $c$. 

The $\ln{\nu}$ growth of the action $\tilde{\mathcal{I}}_{\text{eff}}$ in (\ref{eq:eff_deconf}) at large $\nu/c$ suggests that the original fugacity expansion -- defined at large but finite $c$, has a radius of convergence: 
\be 
\lim_{c\to \infty}\lim_{\nu\to \infty} \left[\sum_{|Y^1|+|Y^2|=\nu} e^{-\tilde{\mathcal{I}}_{\text{eff}}(Y^1,Y^2)}\right]^{-1/\nu}  = 1
\ee
where the sum is over the class of string-like $Y^{1,2}$, and we have assumed that the sum exhibits qualitatively the same scaling behavior of (\ref{eq:eff_deconf}). This is in fact consistent with the existence of the true OPE singularity of virasoro block at $z=1$. It marks the true break down of the fugacity expansion that is not resolved by finite $c$ effects. The effective theory described by (\ref{eq:eff_deconf}) captures this. 

We can in fact make further connections between the explicit expression of (\ref{eq:eff_deconf}) and refined properties for the $z\to 1$ singularity of $\mathcal{V}^t_{vac}(z)$ in the heavy-light limit. Using the fusion kernel of virasoro blocks under crossing, one can derive the leading form of singularity in the heavy-light limit for $\mathcal{V}^t_{vac}(z)$ as \cite{Collier:2018exn}: 
\be \label{eq:HL_UV_sing}
\lim_{z\to 1}\mathcal{V}^t_{vac}(z) \sim (1-z)^{\frac{c}{24}-h_H}=(1-z)^{-\frac{c}{24}\alpha_H^2}
\ee
This singularity can be reproduced if the fugacity expansion coefficients $\mathcal{Z}_\nu$ exhibit the following power-law growth: 
\be\label{eq:ultimate_growth} 
\lim_{c\to \infty}\lim_{\nu \to \infty} \mathcal{Z}_{\nu} \sim \nu^{\frac{c}{24} \alpha_H^2} 
\ee

We now study whether (\ref{eq:ultimate_growth}) can emerge naturally from the effective action (\ref{eq:eff_deconf}): 
\be 
\mathcal{Z}_\nu \sim \sum_{|Y^1|+|Y^2|=\nu} e^{-\tilde{\mathcal{I}}_{\text{eff}}(Y^1,Y^2)}
\ee
To this end, we notice that the effective action at the leading order in large $\nu/c$ and large $c$ depends only on the total number of rows $\bar{n}$, and not on the detailed configurations of the atomic Young tableaux $y^{1,2}_X$. So at this order we can organize the sum over $Y{1,2}$ simply into a sum over $\bar{n}$: \footnote{In going from the sum over $Y^{1,2}$ to $\bar{n}$, there is a degeneracy factor $P(\nu,\bar{n})$ given by the number of partitions of $\nu$ into $\bar{n}$ integers, i.e. $\sum_{|Y^1|+|Y^2|=\nu}(...)\to \sum_{\bar{n}} P(\nu,\bar{n})(...)$. At large $\nu$ we have $P(\nu,\bar{n}) \sim \nu^{\bar{n}}$. This factor does not affect the leading-order analysis in large $\nu/c$ and $c$, so we ignore it.}
\be\label{eq:n_expansion}
\lim_{\nu \gg c\to \infty}\mathcal{Z}_{\nu} \sim  \sum_{\bar{n}} \nu^{\frac{c}{6} \bar{n}(\bar{n}-i\alpha_H)} \times \mathcal{B}_{\bar{n}}(\alpha_H,c)
\ee
The coefficient $\mathcal{B}_{\bar{n}}(\alpha_H,c)$ can be decomposed into the following $\bar{n}$-dimensional integral: 
\bea\label{eq:B_n}
\mathcal{B}_{\bar{n}}(\alpha_H,c)  &=& \sum_{n_1+n_2 = \bar{n}}\int \prod^{n_1}_{i=1} dy^1_i \prod^{n_2}_{i=1} dy^2_i \;\exp{\left[-\frac{c}{6}\tilde{f}(y^1,y^2)\right]}\nonumber\\
&\times &\delta \left(\sum^2_{\alpha =1}\sum^{n_\alpha}_{i=1}y^{\alpha}_i-1\right)
\eea

While it is difficult to compute $\mathcal{B}_{\bar{n}}(\alpha_H,c)$ explicitly using (\ref{eq:B_n}), let us proceed under an ideal scenario in which $\bar{n}$ can be analytically continued into the complex-plane $\bar{n}\in \mathds{C}$. In this case we can re-write the sum over $\bar{n}$ in (\ref{eq:n_expansion}) as a contour integral in $w$: 
\be\label{eq:B_n_2}
\mathcal{Z}_{\nu} \sim \frac{1}{2\pi i} \oint_{\mathcal{C}} \frac{\pi dw}{\sin{(\pi w)}}\; \nu^{\frac{c}{6} w(w-i\alpha_H)}\; \mathcal{B}_{w}(\alpha_H,c)
\ee
This expression is somewhat reminiscent of how the Regge theory for scattering amplitudes emerges from the partial wave decomposition at large $s$:
\be
A(s,t) \sim \sum_{\ell\geq 0} s^{\ell} \cdot f_{\ell}(t)(2\ell+1) 
\ee
and analytically continuing the angular momentum $\ell$. The Regge-behavior is determined by the pole in $\ell$ of $f_\ell(t)$ with the most positive real part. Our expression  (\ref{eq:B_n_2}) differs in that the exponent is a non-trivial function of $w$, and therefore admitting a saddle-point at large $\ln{\nu}$ that is independent of the pole structure of $\mathcal{B}_{w}$. The complex saddle-point $w^*$ can be simply obtained:
\be 
w^* = \frac{i\alpha_H}{2}
\ee
Plugging this back gives precisely: 
\be
\mathcal{Z}_{\nu} \sim \nu^{\frac{c}{24}\alpha_H^2}
\ee
consistent with the prediction (\ref{eq:ultimate_growth}). 

In terms of the instanton gas, the physics of the $z\to 1$ singularity is described again by an instability of instanton proliferation in string-like configurations. This happens already in the $\nu \gg c$ regime, so no finite $c$ corrections can resolve it, thus representing a true instability of the gas. Similar to the low fugacity phase, the proliferation is controlled by a complex saddle-point, this time complexified in the total number of rows $\bar{n}$. On the other hand, the asymmetric between $y^1$ and $y^2$ now disappears at the leading order -- where the only label for the saddle-point configuration is $\bar{n}$. 

We end this section by speculating what might happen when finite $c$ effects are considered in this regime. One possibility is that the run-away direction of the two-dimensional Young tableaux towards larger $\lambda$ might be stabilized at finite $c$, and the two-dimensional Young tableaux may eventually dominate at large $\nu$. This is certainly a reasonable conjecture at $c\sim \mathcal{O}(1)$, since there is no parametric distinctions between the dynamics along the horizontal and vertical directions. If this is true, it might point to a distinct  nature of $z\to 1$ singularity than that of the heavy-light result (\ref{eq:HL_UV_sing}), possibly suggesting some subtleties in the order of limits between $z\to 1$ and $c\to \infty$. We leave this to future investigations.    

\section{Essential singularity at $c=\infty$}\label{sec:large_c}
Having focused on the $z$-dependence of the instanton gas, we make some brief discussion on the $c$-dependence, particularly on the properties of the instanton gas at $c=\infty$. For this purpose, let us write the $c$-dependence explicitly for the fugacity expansion: 
\be\label{eq:fugacity_expansion_c}
\mathcal{Z}^{\text{U(2)}}_{\text{inst}}(c,z) = \sum_\nu \mathcal{Z}_\nu(c) z^\nu 
\ee
The fact that (\ref{eq:fugacity_expansion_c}) has a radius convergence $r(c)$ that is discontinuous in the $c\to \infty$ limit: 
\be
r(c<\infty)=1,\;\;\;r(\infty)=z^*
\ee
means that $\mathcal{Z}^{\text{U(2)}}_{\text{inst}}(c,z)$ is not analytic at $c=\infty$.

On the other hand, we have derived a $c\to \infty$ limit for the instanton gas that is valid in the low fugacity phase for $|z|<z^*$. In more precise terms, this implies that a perturbation series exists: 
\be\label{eq:pert}
\mathcal{Z}^{\text{U(2)}}_{\text{inst}}(c,z) = \mathcal{Z}_0(z)+c^{-1} \mathcal{Z}_1(z)+c^{-2} \mathcal{Z}_2(z)+...
\ee
By deriving the $c$-independent effective action $\mathcal{I}_0$, we are essentially computing the leading-order term in (\ref{eq:pert}). The non-analyticity at $c=\infty$ must therefore come from the series (\ref{eq:pert}) being asymptotic, giving rise to an essential singularity at $c=\infty$. Via the Borel re-summation, a well-defined result for (\ref{eq:pert}) requires adding non-perturbative in $1/c$ contributions and completing it into a trans-series. Resurgence techniques can be used to probe such corrections \cite{Benjamin:2023uib,bissi20241cexpansion2dcfts}. They are explicitly visible at high fugacity $|z|>z^*$, when they dominate and gives a free energy that is $\propto c$, mimicking a deconfined phase. They may also be present in the low fugacity phase $|z|<z^*$ while hidden as exponentially suppressed. Even so, they may still show up through the Stokes' phenomenon if we allow the large central charge $c$ to take values in the complex-plane.  

In the context of 2d CFTs, the $1/c$ corrections can be computed explicitly and systematically using the Zamolodchikov's recursive relation.  Through the AGT correspondence, the non-perturbative effects are encoded in the $1/c$ corrections to $\mathcal{I}_0(Y^1,Y^2)$, which we can study explicitly by expanding (\ref{eq:Nekrasov}) to higher orders in $1/c$. In principle, each order in $1/c$ can be described by an effective theory of instant gas. In the future, it is interesting to study whether a limiting theory emerges at infinite order in $1/c$, in which the mechanism for non-perturbative effects become more transparent. 

\section{Discussions and outlooks}\label{sec:discuss}
In theoretical physics, finding unexpected connections between problems that appear to bear distinct physical contexts is an important way to leapfrog the problems, often by illuminating deep aspects that may otherwise seem obscure. In this paper, we focused on the phenomenon of forbidden singularities that emerge in the heavy-light limit of the vacuum virasoro block. It is a key signature of the more general phenomenon of eigenstate thermalization in 2d holographic CFTs, and shares some common features with the black hole information paradox. Based on properties near the forbidden singularity, we found strong evidence that supports an underlying phase transition in eigenstate thermalized systems. Utilizing the AGT correspondence, which relates properties of virasoro blocks with the physics of $\mathcal{N}=2$ SUSY gauge theories, we found a concrete interpretation for the forbidden singularity in terms of a phase transition regarding the gauge theory instantons. 

In particular, in the equivalence of the heavy-light limit, the instanton gas exhibits a low fugacity phase with a well-defined large $c$ limit. This phase is dominated at large instanton number $\nu$ by gas configurations of string-like Young tableaux with actions $\propto \nu$ -- mimicking a confined phase The forbidden singularity manifests as a critical fugacity for this phase, at which the an instability of instanton proliferation $\nu\to \infty$ occurs. The proliferation is represented by a saddle-point solution for the Young tableaux configuration at large $\nu$. We solve the saddle-point equations explicitly, and found the dominant saddle-point to be complex via analytic continuation. We derive the corresponding critical fugacity, which matches precisely with the forbidden singularity. At finite $c$, the instanton proliferation is indeed a signal for entering a high fugacity phase, which is controlled by configurations with instanton number $\nu \gtrsim c$. We found that at large $\nu/c$ and $c$, the dominant configurations are still those of string-like Young tableaux, but whose actions now $ \propto c\ln{\nu}$ -- mimicking a deconfined phase. Furthermore, we found that such contributions can be re-summed at the leading order in large $\nu/c$ and $c$ to re-produce the $z\to 1$ singular behavior of virasoro blocks derived in the heavy-light limit.   

This paper constitutes a first step towards revealing a much deeper connection between the physics regarding eigenstate thermalization in 2d CFTs and that regarding phase transitions in 4d gauge theories. We end this paper by proposing a few questions for future investigations, so that the nature of this connection can be further elucidated. 

Our results in this paper explicitly examined the low fugacity phase $(z<z^*)$, and the other limit $z\to 1$ of the high fugacity phase. An interesting goal for the future is to zoom into the intermediate regime $z\approx z^*$ that connects both phases. Doing this will allow us to probe deeper into the phase transition, e.g. its universality class, CFT description, etc \cite{WIP}. This can be done via a resolution of the forbidden singularity $z=z_1$. For real fugacity $ z \in \mathds{R}^+$, this can be done in two ways. In the first approach, one begins with large and finite $c$, finds the right scaling behavior of instanton number that dominates in the vicinity of forbidden singularity $z\sim z_1$, e.g. $\nu \sim c $, and studies the effective dynamics by taking the limit of large $c$ -- now as the single large parameter. We expect this regime to also be controlled by stringy Young tableaux. In the second approach, we re-derive the large $c$ effective theory by taking $\epsilon_L= 6h_L/c$ small but finite, as was done in \cite{Wang:2018}. In doing this, the phase transition along the real axis is smoothened into a cross-over, see Figure (\ref{fig:zeros}). In this case, there will not be an sharp transition in the $c$-scaling as (\ref{eq:f_scaling}). The phase transition can then be alternatively studied by tracing the formation of a confining phase in the limit of taking $\epsilon_L \to 0$. In both approaches, we anticipate the analysis to overlap with the those concerning the NS limit of Nekrasov partition functions, with the additional step of taking the heavy-light limit. 

From a broader perspective, we can extend the connection beyond the current context by probing more phenomenon on both sides of the AGT correspondence. For example, it was found that the branch-cuts of first-order transitions (i.e. trajectories of zeros) extends into the Lorentzian sheets in $z$, and crossing them gives rise to the universal hydrodynamic power-law behavior of the late-time correlator in 2d CFTs \cite{Wang:2018,Chen:2017}. Motivated by the connection we found in this paper, it is then interesting to study the physics regarding the nature of the corresponding first order transitions in the 4d gauge theory that occurs by tuning the analytically continued fugacity $z$, or equivalently the gauge coupling $q$. 

On the other hand, while we have associated the low fugacity and high fugacity phases with the ``confined" and ``deconfined" phases, they are motivated by viewing $c$ as an effective gauge group rank, or more broadly the number of local degrees of freedom. This is reasonable for the context of 2d CFTs, but not for the actual 4d gauge theory context. Rather, $c$ enters the 4d gauge theory as a geometric parameter of the $\Omega$-deformation, and somehow acts like an abstract volume for the corresponding instanton gas. In this analogy, the phases are more accurately described as a dilute phase $(z<z^*)$ and a normal phase $(z>z^*)$ of the instanton gas. Despite this, the transition in the behavior of the action (i.e.  microscopic ``free energy") that we observed for the stringy Young tableaux -- changing from $\propto \nu$ at $\nu \ll c$ to $\propto \ln{\nu}$ at $\nu \gg c$, again resembles the dynamical transition of quark potential if $\nu$ can be viewed as the distance between quarks. This suggests that the confinement/deconfinement analogy may bear deeper relevance in the gauge theory context. This is especially the case considering that instanton dynamics, which the phase transition concerns explicitly in this paper, is an important aspect for gauge theory confinement. It is worth exploring this connection in the future. In particular, while in this paper we have worked out the Young tableaux saddle-point dictating the instanton proliferation, its physical implications in terms of the gauge theory dynamics remains obscure. Working out their interpretations, especially how they fit into and interact with the conventional subjects concerning non-perturbative dynamics of $\mathcal{N}=2$ SUSY gauge theories, e.g. the Seiberg-Witten theory, is an important investigations that we leave to the future.   

\section*{Acknowledgment}
We thank Tomoki Nosaka for collaborations at the initial stage of this project. We also thank Hongfei Shu, Futoshi Yagi, Ruidong Zhu for helpful discussions. The work of Y.X, W.S, and H.W is supported by by National Science Foundation of China (NSFC) grant No.\,12175238 and No.\,12447018; the work of C.T is supported by NSFC grant No.\,12475043 and No.\,12447101.  

\appendix 

\section{Effective action $\mathcal{I}_0$ at $c\to \infty$}\label{app:effective_action}

In this appendix we derive the leading large-$c$ effective action for the Young tableaux. Substituting (\ref{eq:AGT_large_c}) into (\ref{eq:Nekrasov}) and keeping only the leading term in each factor, the contribution of a generic box $\Box\in Y^1$ is:
\begin{equation}\label{eq:Y1_generic}
    \mathcal{Z}_{Y^1}(\Box) = \frac{i^2(i-1)(i-\sqrt{1-4\epsilon_H})}{\ell_1 (\ell_1+1)\ell_2(\ell_2-1)}
\end{equation}
Similarly, for a generic box $\Box \in Y^2$, its contribution is given by:
\begin{equation}\label{eq:Y2_generic}
    \mathcal{Z}_{Y^2}(\Box) =\frac{(i-1)^2(i-2)(i-1-\sqrt{1-4\epsilon_H})}{\ell_2 (\ell_2+1)(\ell_1+1)(\ell_1+2)}
\end{equation}
Here by a generic box we mean one whose contribution contains only interior factors, i.e. $\propto c$. Their leading-order $c$-dependence cancel between $\mathcal{Z}_{\text{fund}}$ and $\mathcal{Z}_{\text{vec}}$, leaving the result $c$-independent. Whenever one of the factors in (\ref{eq:Y1_generic}) or (\ref{eq:Y2_generic}) vanishes, we have an exceptional box whose contribution contains at least one boundary factor. The corresponding factor is then given by the next-order $\propto \mathcal{O}(1)$ terms in the large-$c$ expansion. 

The full contribution of a pair $(Y^1,Y^2)$ is obtained by multiplying all boxes. It is convenient to reorganize the product into columns. Denoting by $X_j^i$ the height of the $j$th column of $Y^i$, the contribution of the $j$th column can be written as:
\bea\label{eq:column_total}
\mathcal{Z}_j(X^1_j,X^2_j) &=& \mathcal{Z}_{Y^1}(X^1_j,X^2_j)\times \mathcal{Z}_{Y^2}(X^1_j,X^2_j)\nonumber\\
&=& \mathcal{Z}_1(X^1_j) \times \mathcal{Z}_2(X^2_j)\times \mathcal{Z}_{\partial}(j)
\eea

The first two factors in \eqref{eq:column_total} are the contributions from the interior factors (\ref{eq:Y1_generic},\ref{eq:Y2_generic}), which depend only on the vertical positions of the boxes. So upon multiplying all the interior factors in a column, the result only depend on the column lengths $X^{1,2}_j$. It may appear that due to interactions between $Y^1$ and $Y^2$, the result should also depend on $\Delta X_j = |X^2_j-X^1_j|$. It turns out that such dependence cancel out in a remarkable way. The final result takes the simple form:
\be\label{eq:column_bulk}
\mathcal{Z}_{\alpha}(X) = \frac{\Gamma(X+2-\alpha-i\alpha_H)}{\Gamma(X+3-\alpha)\Gamma(2-\alpha-i\alpha_H)},\;\alpha=1,2
\ee
Notice that in addition to those from (\ref{eq:Y1_generic}) and (\ref{eq:Y2_generic}), (\ref{eq:column_bulk}) also includes contributions from the interior factors associated with the exceptional boxes, whose $c$-scalings do not necessarily cancel between $\mathcal{Z}_{\text{fund}}$ and $\mathcal{Z}_{\text{vec}}$ due to the boundary factors. Fortunately, a counting analysis shows that the total number of boundary factors equal between $\mathcal{Z}_{\text{fund}}$ and $\mathcal{Z}_{\text{vec}}$, see Appendix (\ref{app:cancellation}). The result (\ref{eq:column_bulk}) therefore remains $c$-independent.

The factor $\mathcal{Z}_{\partial}(j)$ multiplies all the boundary factors in the $j$-th column of both $Y^1$ and $Y^2$, which are explicitly $c$-independent at the leading order in large $c$ expansion. Unlike the contributions from the interior factors, these contributions depend on the horizontal positions of the boxes. Multiplying them thus gives rise to dependence on the row-lengths $Y^{1,2}_i$. In more details, the boundary factors are organized into three classes --- the upper edges, the lower edges, and the immersed lower edges. We give their identifications and contributions below.
\paragraph{Upper edges.}
These arise from the first row of $Y^1$ and the first two rows of $Y^2$, and contribute as numerator factors via $\mathcal{Z}_{\text{fund}}$:
\begin{itemize}
\item $(i=1,j)\in Y^1:\;\;\;(j-1-\Delta+2h_L)$ 
\item $(i=1,j)\in Y^2:\;\;\;(j-1+\Delta)^2$
\item $(i=2,j)\in Y^2:\;\;\;(j-2+\Delta +2h_L)$
\end{itemize}

\paragraph{Lower edges.}
These arise from the bottom boxes of the two columns and contribute as denominator factors via $\mathcal{Z}_{\text{vec}}$:
\begin{itemize}
\item $(i=X^1_j,j)\in Y^1:\;\;\;-(Y^{1}_{[X^{1}_j]}-j+1)$
\item $(i=X^2_j,j)\in Y^2:\;\;\;-(Y^{2}_{[X^{2}_j]}-j+1)$
\end{itemize}

\paragraph{Immersed lower edges.}
These arise from the off-diagonal cases $\alpha\neq \beta$, in which the lower edge of one Young diagram is ``immersed" in the other. They contribute as additional denominator factors via $\mathcal{Z}_{\text{vec}}$:
\begin{itemize}
\item $(i=X^2_j-1,j)\in Y^1:\;\;\;-(Y^{1}_{[X^2_j-1]}-j+2-2\Delta)$
\item $(i=X^2_j,j)\in Y^1:\;\;\;-(Y^{1}_{[X^2_j]}-j +1-2\Delta)$
\item $(i=X^1_j+1,j)\in Y^2:\;\;\;-(Y^{2}_{[X^1_j+1]}-j+2\Delta)$
\item $(i=X^1_j+2,j)\in Y^2:\;\;\;-(Y^{2}_{[X^1_j+2]}-j-1+2\Delta)$
\end{itemize}
The factor $\mathcal{Z}_{\partial}(j)$ is then obtained by multiplying all such factors if they exist. 

The rest of the computation is tedious -- having to multiply all columns and keep track of various cases, but straightforward. \footnote{A minor subtlety arises when the two boundaries touch, \emph{i.e.} $X_j^2=X_j^1+1$, in which case an additional factor of $(-1)$ is produced and should be included in $\mathcal{Z}_{\partial}(j)$.} We simply write down the result:
\bea
\prod_{j} \mathcal{Z}_j(X^1_j,X^2_j)=e^{-\mathcal{I}_0(Y^1,Y^2)}
\eea
as a total action $\mathcal{I}_0$ consisting of four terms: 
\be
\mathcal{I}_0(Y^1,Y^2)=\mathcal{I}_1(Y^1)+\mathcal{I}_2(Y^2)+\mathcal{I}_{\text{int}}(Y^1,Y^2)+i\Theta
\ee
The ``self-energy" terms $\mathcal{I}_{1,2}$ are given by:
\bea
&\mathcal{I}_{1}(Y^1)& = \sum_X N^1_X \ln{\left[\frac{\Gamma(X+2)\Gamma(1-i\alpha_H)}{\Gamma(X+1-i\alpha_H)}\right]}\nonumber\\
&+& \ln{\left[\frac{\Gamma(2h_L-\Delta)}{\Gamma\left(Y^1_1+2h_L-\Delta\right)}\right]}+\sum_{X} \ln{\Gamma(N^1_X+1)}  \nonumber\\
&\mathcal{I}_{2}(Y^2)& = \sum_X N^2_X \ln{\left[\frac{\Gamma(X+1)\Gamma(-i\alpha_H)}{\Gamma(X-i\alpha_H)}\right]}\nonumber\\
&+&2\ln{\left[\frac{\Gamma(\Delta)}{\Gamma\left(Y^2_1+\Delta\right)}\right]}+\ln{\left[\frac{\Gamma(2h_L+\Delta-1)}{\Gamma\left(Y^2_2+2h_L+\Delta-1\right)}\right]}\nonumber\\
&+& \sum_{X} \ln{\Gamma(N^2_X+1)} 
\eea
where we have defined the ``occupation numbers" $N^{1,2}_X$ with respect to $Y^{1,2}_X$ as:
\be
N^{1,2}_X = Y^{1,2}_{X}-Y^{1,2}_{X+1},\;\;\;\;Y^{1,2}_X = \sum_{X'\geq X} N^{1,2}_{X'}
\ee
The interaction term $\mathcal{I}_{\text{int}}$ is given by:
\bea
&&\mathcal{I}_{\text{int}}(Y^1,Y^2) = \sum_{X} \ln\Bigg[\frac{\Gamma\left((Y^1_X-Y^2_{X+1})^++1-2\Delta\right)}{\Gamma\left((Y^1_X-Y^{2}_{X})^++1-2\Delta\right)} \nonumber\\
&\times & \frac{\Gamma\left((Y^1_X-Y^2_{X+2})^++2-2\Delta\right)\Gamma\left((Y^2_{X}-Y^1_{X})^++2\Delta\right)}{\Gamma\left((Y^1_X-Y^{2}_{X+1})^++2-2\Delta\right)\Gamma\left((Y^2_{X}-Y^{1}_{X-1})^++2\Delta\right)}\nonumber\\
&\times &\frac{\Gamma\left((Y^2_X-Y^1_{X-1})^+-1+2\Delta\right)}{\Gamma\left((Y^2_X-Y^{1}_{X-2})^+-1+2\Delta\right)}\Bigg]
\eea
We have absorbed all factors of $(-1)$ into a phase index $\Theta$ that is given via a counting analysis by:
\bea
&&\Theta =Y^1_1-Y^2_2+(Y^2_1-\max\{Y^1_1,Y^2_2\})^+\nonumber\\
&+& \sum_{X}(\min\{Y^1_X,Y^2_{X+1}\}-\max\{Y^1_{X+1},Y^2_{X+2}\})^+
\eea
In writing these expressions, we have also defined:
\be
A^+ = \begin{cases}
A,\;\;\;A>0\\
0,\;\;\;\;A\leq 0
\end{cases}
\ee

\section{Cancellation of $c$-dependence among exceptional boxes}\label{app:cancellation}
A key property of the effective action $\mathcal{I}_0$ is its $c$-independence. In this appendix, we lay out the underlying mechanism that manages the cancellation at the leading order in large $c$. 

As was discussed, the only source of non-cancellation stems from exceptional cases to the expressions (\ref{eq:Y1_generic}) and (\ref{eq:Y2_generic}), i.e. cases when a factor in the numerator or denominator vanishes, contributing a factor of $c^{-1}$ or $c$ respectively. Define the occurrence number for a vanishing factor in the numerator by $(1)$ and that of the denominator by $(-1)$, we show that the occurrences cancel between the numerator and denominator, thereby resulting in the overall $c$-independence. From (\ref{eq:Y1_generic}) and (\ref{eq:Y2_generic}), we identify the following cases where a factor from the numerators vanishes: 
\begin{itemize}
\item P1: $\;i=1,\;\;\Box=(i,j)\in Y^1,\;\;X^{1}_j\geq 1$
\item P2$\times 2$: $\;i=1,\;\;\Box=(i,j)\in Y^2,\;\;X^{2}_j \geq 1$
\item P3: $\;i=2,\;\;\Box=(i,j)\in Y^2,\;\;X^2_j\geq 2$
\end{itemize}
and the following cases when a factor in the denominator vanishes: 
\begin{itemize}
\item Q1: $\;\ell_{Y^\alpha}(\Box)=0,\;\;\Box \in Y^{\alpha},\;\;X^\alpha_j\geq 1$
\item Q2: $\;\ell_{Y^2}(\Box)=0,\;\;\Box \in Y^1,\;\;X^1_j\geq X^2_j\geq 1$
\item Q3: $\;\ell_{Y^2}(\Box)=1,\;\;\Box \in Y^1,\;\;X^1_j\geq X^2_j-1\geq 1$
\item Q4: $\;\ell_{Y^1}(\Box)=-2,\;\;\Box \in Y^2,\;\;X^2_j\geq X^1_j+2$
\item Q5: $\;\ell_{Y^1}(\Box)=-1,\;\;\Box \in Y^2,\;\;X^2_j\geq X^1_j+1$
\end{itemize}
where we have also indicated the condition for the case to occur. Notice that the case P2 occurs with degeneracy 2, i.e. two factors vanish when it happens. We can restrict to a particular column index $j$, and compute the total occurrence $N$ for the $j$-th column of $Y^{1,2}$ by adding up the signed Boolean functions representing the conditions for P1-P3 and Q1-Q5: 
\bea\label{eq:total_N}
&&N=\mathcal{B}(X^1_j \geq 1)+2\mathcal{B}(X^2_j\geq 1)+\mathcal{B}(X^2_j\geq 2)\nonumber\\
&-&\sum^2_{\alpha=1}\mathcal{B}(X^\alpha_j\geq 1)-\mathcal{B}(X^1_j\geq X^2_j\geq 1)-\mathcal{B}(X^2_j\geq X^1_j+2)\nonumber\\
&-& \mathcal{B}(X^1_j\geq X^2_j-1\geq 1)-\mathcal{B}(X^2_j \geq X^1_j+1)
\eea
This can be simplified by deriving the Boolean identities:
\bea
&& \mathcal{B}(X^1_j\geq X^2_j\geq 1)+\mathcal{B}(X^2_j\geq X^1_j+2)\nonumber\\
&\equiv & \mathcal{B}(X^2_j \geq 1) -\mathcal{B}(X^2_j=X^1_j+1)
\eea
and
\bea
&&\mathcal{B}(X^1_j\geq X^2_j-1\geq 1)+\mathcal{B}(X^2_j\geq X^1_j+1)\nonumber\\
&\equiv & \mathcal{B}(X^2_j \geq 2) +\mathcal{B}(X^2_j=X^1_j+1)
\eea
Plugging these identities into (\ref{eq:total_N}), we obtain that $N=0$. The cancellation is valid for each column index $j$, and thus is valid for the total contribution.  

We emphasize that the heavy-light arrangement of the external operators plays a crucial role in this cancellation. The vanishing conditions P1-P3 rely on the specific large $c$ assignments in (\ref{eq:AGT_large_c}), in particular $\mu_2\approx 3c/2$. For example, had we chosen to take the all-heavy limit, some of the vanishing cases in P1-P3 will not be true, and the cancellation will fail. 

\section{General reduced-form of $\tilde{\Theta}$}\label{app:phase_property}
In this appendix, we prove the general property that for any cross-pattern between $y^1$ and $y^2$, the expression (\ref{eq:Theta_rescaled}) of $\tilde{\Theta}$ reproduced below: 
\bea\label{eq:Theta_rescaled_2}
\tilde{\Theta}&=&2y^1_1+2y^2_0+(y^2_0-\max\{y^1_1,y^2_1\})^+\nonumber\\
&+&\sum_{X=1}^{\infty}(\min\{y^1_X,y^2_X\}-\max\{y^1_{X+1},y^2_{X+1}\})^+\nonumber\\
&-&\sum_{X=0}^{\infty}(\Delta^{2,1}_X)^++\sum_{X=0}^{\infty}(-\Delta^{1,2}_{X+1})^+ - \sum^\infty_{X=0} \Delta^{2,1}_X
\eea
can be reduced to the form of (\ref{eq:phase_property}) reproduced below: 
\be\label{eq:phase_property_2}
\tilde{\Theta} = \sum^\infty_{X=1} c_X\cdot y^{1}_X + \sum^\infty_{X=0} d_X \cdot y^2_X
\ee
with $c_X$ and $d_X$ are odd integers. 

To proceed, we separate (\ref{eq:Theta_rescaled_2}) into the ``base-line" terms including the first two terms and the last term -- they do not depend on cross-patterns, and the remaining correction terms:
\be
\tilde{\Theta} = 3y^1_1+y^2_0-\sum^\infty_{X=1}y^2_X+\sum^\infty_{X=2}y^1_X + \text{corrections}
\ee
The base-line terms are indeed of the form (\ref{eq:phase_property_2}). All we need to prove is that the correction terms modify the coefficients only by even integers. 

To this end, we fix any particular $y^{1,2}_X$ and examine how the correction terms modify its coefficient in $\tilde{\Theta}$. To be explicit, we focus on the case of $y^1_{X\geq 2}$. The modification to its coefficient from the correction terms can be expressed in terms of the Boolean functions: 
\bea
\Delta c_X &=& \mathcal{B}(y^2_{X}>y^1_X>y^2_{X+1}) - \mathcal{B}(y^2_{X-1}>y^1_X>y^2_X)\nonumber\\
&+&\mathcal{B}(y^2_{X-1}>y^1_X)- \mathcal{B}(y^2_{X+1}>y^1_X)
\eea
Using basic Boolean identities of the form: 
\be
\mathcal{B}(a)+\mathcal{B}(\lnot{a})=1,\; \mathcal{B}(a\& b) = \mathcal{B}(a)\cdot \mathcal{B}(b)
\ee 
we can reduce it further:
\bea
\Delta c_X &=& \mathcal{B}(y^2_X>y^1_X)+\mathcal{B}(y^1_X>y^2_X)\mathcal{B}(y^2_{X+1}>y^1_X)\nonumber\\
&+& \mathcal{B}(y^2_{X-1}>y^1_X)\mathcal{B}(y^2_X>y^1_X)-2\mathcal{B}(y^2_{X+1}>y^1_X)\nonumber\\
&=& 2\mathcal{B}(y^2_X>y^1_X)-2\mathcal{B}(y^2_{X+1}>y^1_X)
\eea
To see the last equality, we firstly notice that the product term in the first line vanishes identically -- the two conditions are not compatible with $y^2_X>y^2_{X+1}$; one can also deduce that $\mathcal{B}(y^2_X>y^1_X)\mathcal{B}(y^2_{X-1}>y^1_X)=\mathcal{B}(y^2_{X}>y^1_X)$, since logically we have $y^2_X >y^1_X\to y^2_{X-1}>y^1_X$. This proves that $\Delta c_X$ is always an even integer, thus $c_X = 1+\Delta c_X$ is always an odd integer as claimed. 

The proof for other terms, including $y^2_{X\geq 2}$ and the boundary cases $\lbrace y^1_1,y^2_0,y^2_1\rbrace$, proceeds in completely analogous manners. 

\section{Effective action for $\nu\gg c \gg 1$}\label{app:large_nu}
In this appendix, we evaluate the instanton partition function 
$\mathcal{Z}(Y^1,Y^2)$ in the regime of $\nu\gg c\gg 1$. To decouple finite $c$ corrections in this regime, we assume that $\nu/c$ is fixed at some large ratio while sending $c\to \infty$. In this limit we can focus on the string-like
Young tableaux of the scaling: $Y^\alpha_X = \nu\,y^\alpha_X,\;\;X=1,...,n_\alpha$. The numbers of rows $n_\alpha$ are treated as part of the variational data. We derive the effective action $\mathcal{Z}(Y^1,Y^2)$ in this regime as a function of $(y^1,y^2)$.
We organize the computation into rows -- by first summing over boxes in each row and then summing results over the row index. We do this separately for the numerator ($\mathcal{Z}_{\text{fund}}$) and denominator ($\mathcal{Z}_{\text{vec}}$). Each box $\Box$ in $Y^\alpha$ contributes $+4$ factors from
$\mathcal{Z}_{\text{fund}}$ (product over the four masses $\mu_k$)
and $-4$ factors from $\mathcal{Z}_{\text{vec}}$ (product over
$\beta=1,2$). 

\paragraph{\underline{$\ln{\mathcal{Z}_{\text{fund}}}$ contribution.}}
The basic ingredients for computing $\ln{\mathcal{Z}_{\text{fund}}}$ consists of the following summations over a particular row $X$ with length $Y_X$:
\begin{eqnarray}\label{eq:ap_S1}
    \sum_{j=1}^{Y_X}\ln(b_X c+6j)
    &=& Y_X\ln 6 + \ln\Gamma\left(Y_X+\frac{b_X}{6} c+1\right)\nonumber\\
    &-& \ln\Gamma\left(\frac{b_X}{6} c+1\right).
\end{eqnarray}
There are four such sums for each row corresponding to $\mu_{i=1,...,4}$. Depending on which Young diagram it comes from, the coefficient $b_X$ takes the following values at row $X$:
\begin{equation}\label{eq:chi_table}
\left[
\begin{array}{c|cccc}
& \mu_1 & \mu_2 & \mu_3 & \mu_4\\\hline
Y^1 & X\;\;\; & X-1 \;\;\;& X\;\;\; & X-i\alpha_H \\[2pt]
Y^2 & X-1\;\; & X-2\;\;\; & X-1\;\;\; & X-1-i\alpha_H
\end{array}
\right]
\end{equation}
We apply the Stirling expansion of (\ref{eq:ap_S1}) for $Y_X\gg c$. The leading order terms in this approximation are $Y_X\ln{Y_X}-Y_X$, which are identical for all sums of the form (\ref{eq:ap_S1}) as well as those (\ref{eq:ap_S2}) from $\ln{\mathcal{Z}}_{\text{vec}}$. They cancel out and will be neglected in subsequent analysis. The surviving terms of $\ln\mathcal{Z}(Y^1,Y^2)$ begin at the order $c \ln{(\nu/c)}$. 
Summing over the row-index $X$ as well as the four masses, we obtain that: 
\begin{eqnarray}\label{eq:fund_large_nu}
 &\ln{\mathcal{Z}_{\text{fund}}} & \sim  \sum^{2}_{\alpha=1}\sum^{n_\alpha}_{X=1} \frac{A^\alpha_X c}{6}\,\left(\ln Y^\alpha_X\right) \nonumber\\
 &\sim & \frac{c}{6}\left(2n_1^2+2n_2^2+n_1-3n_2-i \bar{n}\alpha_H\right) \ln{\left(\frac{\nu}{c}\right)}\nonumber\\
 &+& \sum^{2}_{\alpha=1}\sum^{n_\alpha}_{X=1} \frac{A^\alpha_X c}{6}\,\left(\ln y^\alpha_X \right)
\end{eqnarray}
where $\bar{n}=n_1+n_2$  is the total number of rows, and $A^\alpha_X$ is given by: 
\be
A^\alpha_X = 4X-1-i\alpha_H-4\delta_{\alpha,2}
\ee

\paragraph{\underline{$\ln{\mathcal{Z}_{\text{vec}}}$ contribution.}}
The computation of $\ln{\mathcal{Z}_{\text{vec}}}$ is slightly more involved. In summing over boxes in a row $X$ with length $Y_X$, one encounters the following ingredients: 
\be \label{eq:ap_S2}
\sum^{Y_X-1}_{a=0} \ln{\left(-g_X(a) c + 6a\right)}
\ee
where we remind $a_X = Y_X-j$ is the arm-length of the box. Unlike those of $\ln{\mathcal{Z}_{\text{fund}}}$, this cannot be summed into a closed-form expression because $g_X(a)$ is a piecewise constant function of $a$. There are 4 such sums for each row. Depending on which Young diagram it comes from, the functions $g_X(a)$ take the following form in terms of the leg-lengths $\ell$ for the four choices of $\alpha,\beta\in \lbrace1,2\rbrace$ in (\ref{eq:Nekrasov}):  
\begin{equation}
\label{eq:g_table}
\left[
\begin{array}{c|cccc}
\\\hline
Y^1 & \ell_{Y^1}\;\;\; & \ell_{Y^1}+1 \;\;\;& \ell_{Y^2}-1\;\;\; & \ell_{Y^2} \\[2pt]
Y^2 & \ell_{Y^2}\;\; & \ell_{Y^2}+1\;\;\; & \ell_{Y^1}+1\;\;\; & \ell_{Y^1}+2
\end{array}
\right]
\end{equation}
It is easy to see that the contributions from $a\lesssim c$ are universal, i.e. independent of $y^{1,2}_X$. They enter as a constant to the effective action. The dynamical part of the effective action comes from summing over $a\gg c$, which can be approximated as: 
\be
\sum^{Y_X-1}_{a=0} \ln{\left(-g_X(a) c + 6a\right)} \approx  -\frac{c}{6}\sum^{Y_X}_{a\gtrsim c} \frac{g_X(a)}{a} +...
\ee
where ... denotes terms that either cancel out against other contributions (e.g. $Y_X\ln Y_X$) or are sub-leading in large $\nu/c$ and large $c$. Since $a \gg g_X(a)$ we can further approximate the summation into a sum of logs: 
\bea\label{eq:log_approx}
\sum^{Y_X}_{a\gtrsim c} \frac{g_X(a)}{a} &\approx & g_X(0)\ln{(\tilde{a}_1/c)}+ g_X(\tilde{a}_1) \ln{\left(\tilde{a}_2/\tilde{a}_1\right)}\nonumber\\
&+& g_X(\tilde{a}_2) \ln{\left(\tilde{a}_3/\tilde{a}_2\right)}+...
\eea
where $\tilde{a}_n$ are points where the piecewise function $g_X(a)$ jumps. In particular, for the diagonal cases, i.e. first two columns of (\ref{eq:g_table}), we have $\tilde{a}_n=Y_X-Y_{X+n}$; for the off-diagonal cases, i.e. the last two columns of (\ref{eq:g_table}), we have $\tilde{a}^{\alpha}_n = Y^{\alpha}_X - Y^{\bar{\alpha}}_{s^\alpha_X+n}$, where the index $s^\alpha_X$ is defined as:
\be
s^{\alpha}_X \equiv \text{Max}\lbrace k, y^{\bar{\alpha}}_k > y^\alpha_X\rbrace = \sum^{n_{\bar{\alpha}}}_{k = 1} \Theta(y^{\bar{\alpha}}_k>y^\alpha_X)
\ee
where $\bar{\alpha}$ is defined such that $\alpha \neq \bar{\alpha}$. 

Neglecting terms that cancel out, the leading order contribution of (\ref{eq:ap_S2}) is also of order $c\ln{(\nu}/c)$ and is only given by the first term $\sim (c/6)\cdot g_X(0) \ln{(\nu/c)}$ in (\ref{eq:log_approx}). Summing over (\ref{eq:ap_S2}) for all rows using (\ref{eq:log_approx}), we can obtain the following estimate for $\ln\mathcal{Z}_{\text{vec}}$: \footnote{In deriving this, we have assumed that the piecewise integer function $g_i(a_i)$ only jumps by 1 across discontinuities. This is valid for generic non-identical real numbers $\lbrace y^\alpha_i \rbrace$. } 
\bea\label{eq:vec_large_nu}
&&\ln\mathcal{Z}_{\text{vec}} \sim \frac{c}{6}\left(n_1^2+n_2^2-3 n_2+n_1- 2\sum^2_{\alpha=1} \sum^{n_\alpha}_{X=1} s^\alpha_X\right) \ln{\left(\frac{\nu}{c}\right)} \nonumber\\
&+&\frac{c}{3}\sum^{2}_{\alpha=1}\sum^{n_\alpha}_{X=1}\Bigg(\sum^{n_\alpha}_{k=X+1}\ln{\left(y^\alpha_X-y^\alpha_k\right)}+\sum^{n_{\bar{\alpha}}}_{k=s^\alpha_X+1}\ln{\left(y^\alpha_X-y^{\bar{\alpha}}_{k}\right)}\nonumber\\
&-&\left(\bar{n}-2X+2\delta_{\alpha,2}\right) \ln{y^\alpha_X}\Bigg)
\eea
Similar to before, in writing this we have formally defined $y^{\alpha}_{i}=0$ for $i> n_{\alpha}$.

Now we can sum over the contributions (\ref{eq:fund_large_nu},\;\ref{eq:vec_large_nu}) from $\ln{\mathcal{Z}_{\text{fund}}}$ and $\ln{\mathcal{Z}_{\text{vec}}}$. Some simplifications occur along the way and we derive the following leading order effective action in the limit $\nu \gg c\gg 1$: 
\bea\label{eq:eff_deconf_1}
\tilde{\mathcal{I}}_{\text{eff}}\left(\nu \cdot y^1,\nu \cdot y^2\right) &=&  \frac{c}{6}\ln{\left(\frac{\nu}{c}\right)}\bar{n}(i\alpha_H-\bar{n})\nonumber\\
&+& \frac{c}{6} \tilde{f}\left(y^1,y^2\right) + ... 
\eea
The shape-dependence, entering at sub-leading order in large $\nu/c$, is given by $\tilde{f}(y^1,y^2)$ at the leading order in large $c$:
\bea\label{eq:eff_deconf_2}
\tilde{f}\left(y^1,y^2\right) &=& \sum^{2}_{\alpha=1}\sum^{n_\alpha}_{X=1}\Bigg(\sum^{n_\alpha}_{k>X}\ln{\left(y^\alpha_X-y^\alpha_k\right)^2} + \sum^{n_{\bar{\alpha}}}_{k>s^\alpha_X}\ln{\left(y^\alpha_X-y^{\bar{\alpha}}_{k}\right)^2}\nonumber\\
&-& \left(2\bar{n}-i\alpha_H-1\right) \ln{y^\alpha_X}\Bigg)
\eea
We mention that the sum $\sum_{\alpha=1}^{2}\sum^{n_\alpha}_{X=1} s^\alpha_X$ in (\ref{eq:vec_large_nu}) simplifies due to the following identity:
\bea
\sum_{\alpha=1}^{2}\sum^{n_\alpha}_{X=1} s^\alpha_X &=& \sum^{n_\alpha}_{X=1} \sum^{n_{\bar{\alpha}}}_{Z=1} \left(\Theta(y^\alpha_X<y^{\bar{\alpha}}_Z)+\Theta(y^{\bar{\alpha}}_Z>y^\alpha_X)\right)\nonumber\\
&=& n_1 n_2
\eea

\bibliography{ref.bib}% Produces the bibliography via BibTeX.
\end{document}